\documentclass[preprint,12pt]{elsarticle}

\usepackage{amsmath,amssymb}
\usepackage{bm}
\usepackage{graphicx}
\usepackage{longtable}
\usepackage{multirow}

\journal{Nuclear Physics A}

\begin{document}

\begin{frontmatter}

\title{Effective interaction dependence of the liquid-gas phase transition in symmetric nuclear matter}
\date{\today}

\author{Arnau Rios}
\address{Department of Physics, Faculty of Engineering and Physical Sciences, University of Surrey, 
Guildford, Surrey GU2 7XH, United Kingdom}
\ead{a.rios@surrey.ac.uk}

\begin{abstract}
The liquid-gas phase transition for homogeneous symmetric nuclear matter is studied in the mean-field approximation. Critical properties are computed using a comprehensive group of Skyrme and Gogny forces in an effort to elucidate the effective interaction dependence of the results. Analytical models for the thermodynamical and critical properties are discussed and compared to an extensive set of mean-field data. In agreement with these models, a tight correlation is found between the flashing and the critical points. Accurate predictions for the critical temperature, based on saturation properties, can only be obtained after the density dependence of the effective mass is properly taken into account. While the thermodynamical properties coming from different mean-fields do not follow a law of corresponding states, the critical exponents for all the mean-fields have been found to be the same. Their values coincide with those predicted by the Landau mean-field theory of critical phenomena. 
\end{abstract}


\end{frontmatter}

\section{Introduction}
\label{sec:intro}

The liquid-gas phase transition is the canonical example of a phase transition in statistical physics \cite{huang}. At a very basic level, the existence of the transition can be directly linked to the structure of the two-body interaction \cite{pathria}. The inter-atomic potential is repulsive at short relative distances and attractive at intermediate and long distances. At low temperatures, atoms tend to sit at a preferred relative distance from each other, which favors the formation of a liquid phase. When temperature is increased, however, the thermal motion of the particles is activated and the repulsive, short-range structure of the force is probed. The formation of a gas phase is therefore facilitated until, above a certain critical temperature, $T_c$, the system completely evaporates \cite{pathria}. 

In principle, the classical picture of the liquid-gas phase transition also applies to nuclear systems, for which the interaction in the central channel has a similar structure \cite{siemens83,ring}. Critical properties might therefore be able to provide information on the underlying nucleon-nucleon (NN) interaction. In practice, however, the existence of this transition is difficult to establish unambiguously \cite{pochodzalla97}. First of all, nuclei are finite systems and the empirical signatures of the phase transition might be washed out by their finite size effects \cite{gulminelli99}. Moreover, it is difficult, if not impossible, to reconstruct the dynamics of hot nuclei from the available experimental measurements and thus the conditions at which these reactions happen cannot be easily determined \cite{tsang06}. An example of these difficulties is provided by the plateau in the nuclear caloric curve \cite{pochodzalla97}. This phenomenon, observed by different experimental groups with various reactions in a relatively wide collision energy range \cite{natowitz02}, is often taken as a faithful demonstration of the occurrence of the liquid-gas phase transition \cite{pochodzalla95}. Nevertheless, alternative explanations for the appearance of such a plateau, that do not resort to any phase transition, can also be found in the literature \cite{sobotka04,shetty09}. 

Nowadays, experimental efforts are focusing on the study of hot, isospin asymmetric nuclear systems  \cite{tsang01,sfienti09}. Phenomena such as isoscaling have fueled the search for connections between the critical point of asymmetric nuclei and the isospin dependence of the equation of state. Theoretical analysis of these reactions could provide additional constraints to the isospin dependence of nuclear forces, besides those given by giant resonances \cite{piekarewicz07} or transport models \cite{danielewicz02}. An effort in this direction, linking the symmetry properties to isoscaling data with a Skyrme-type input for statistical multi-fragmentation models, is already underway \cite{souza09}. Statistical multi-fragmentation models are somewhat phenomenological, but provide excellent descriptions of several observables, such as yields \cite{tsang06}. 

Less connected to reactions themselves but closer to the spirit of nuclear structure studies, there have been several attempts to generalize zero temperature calculations to the finite temperature domain. Bulk properties of excited nuclear matter can be computed in such approaches, aiming to describe the freeze-out conditions and the creation of the primary fragments in nuclear collisions. Liquid drop formulae have been used extensively in this context \cite{levit85,song91}, in spite of the uncertainties associated with the phenomenological finite temperature dependence of their surface and Coulomb terms \cite{ravenhall83}. The calculations for isospin asymmetric drops are among the few that are able to predict isotopic shifts of the critical properties \cite{zhang96,li04}, which are of interest given the available experimental data \cite{sfienti09}. 

A more controlled and consistent approach to finite temperature nuclear properties can be obtained within the self-consistent mean-field framework \cite{bender03}. Mean-field calculations at finite temperature include asymmetry, surface and Coulomb effects at once, with no need of additional phenomenological parameterizations \cite{bonche84}. They are based on sound theoretical grounds (statistical mechanics together with quantum many-body theory) \cite{vautherin96} and yield predictions that are equally valid at zero and finite temperature. Moreover, at a formal level, the self-consistent mean-field approach is closely connected to density functional theory \cite{bender03}. The equations and calculations presented in the following can also be obtained within the finite temperature extension of density functional theory \cite{mermin65,gupta80}. The latter theory is among the few available frameworks that can provide a controlled methodology to relate zero- and finite-temperature predictions, both for nuclei and nuclear matter. So far, calculations for hot nuclei have been generally performed on selected spherical isotopes, mostly aimed at understanding the structural changes induced by temperature \cite{brack74,bonche85,sil04}. In principle, finite temperature mean-field calculations could also be used to find the limiting temperatures for selected isotopic chains. The necessary extension to deformed isotopes should be feasible with present computational capabilities. Yet, in spite of their great interest, systematic studies of this type for a relevant range of isotopes are still lacking. 

Before such a program is implemented, however, the model dependence of the calculations should be assessed. Key among the possible sources of uncertainty is the dependence associated with the underlying NN effective interaction. It is well-known that different effective interactions predict different bulk properties for nuclear matter \cite{stone03}. This statement also applies at finite temperature and, as a result, different liquid-gas critical points are found for different forces. In theory, a reliable determination of the critical point could be helpful in constraining the properties of the effective interaction \cite{natowitz02b}. In this contribution, I will discuss the case of symmetric nuclear matter, where this connection should be more straightforward due to the absence of symmetry, Coulomb and surface terms. 

Because of its simplified nature, the mean-field picture is presumably unable to describe correctly the whole phase diagram of nuclei or nuclear matter. The level density lacks the effect of complex many-body excitations and therefore underestimates that of nuclear systems \cite{chomaz04}. Even in infinite matter within a fully quantal description, the effect of beyond mean-field correlations and clusterization on the density of states and the thermodynamics of the system is difficult to quantify. As for finite nuclei, clusterization is completely absent in the mean-field picture, but it is known to play a major role in hot nuclear systems \cite{typel10}. In general, the mean-field approximation provides a poor description of the critical properties of finite, charged systems \cite{gulminelli99}. Related to that, information gathered in time-dependent studies indicates that the mean-field description gives a particularly poor description of the spinodal zone \cite{belkacem94,ono07}. The fragmentation mechanism of nuclei is sensitive to many-body correlations and dynamical time-scales in the reaction dynamics, and the mean-field picture fails in describing such effects. 

Consequently, the mean-field description of nuclear matter should not be taken as meaningful description of the liquid-gas transition in nuclear systems. As a matter of fact, the study presented here does not try to provide a quantitatively accurate picture for such critical properties. Instead, the aim of this study is to investigate the effective interaction dependence of the liquid-gas transition properties. As such, this can be taken as a first methodological contribution to test what information, if any, can be obtained from such analysis. Homogeneous nuclear matter within a mean-field picture provides the simplest testing ground for such an analysis due also to the large number of available effective interactions. Other theoretical approaches, which, for instance, provide a more realistic description of the spinodal zone, will have an underlying dependence on the effective interaction and similar analysis can be carried out for these cases. 

In addition to providing a methodological test, the study of the critical properties of homogeneous matter is also of theoretical interest in connection with recent developments. First of all, the liquid-gas phase transition can nowadays be computed from many-body approaches that go beyond the mean-field. Starting from realistic phase-shift equivalent NN interactions and using finite temperature many-body techniques to find the thermodynamical properties, the liquid-gas phase transition can be constructed \cite{baldo04}. Such \emph{ab initio} calculations provide a relatively wide range of predictions for the critical properties depending on the underlying interaction and many-body approximation that is used \cite{rios08,soma09}. It is therefore important to investigate whether a similar wide range of values is obtained within mean-field theory and whether such an interaction-dependence of the results can be eliminated.

Secondly, the phase transition for homogeneous matter has important connections to astrophysics, particularly to the onset of instabilities in the isospin asymmetric nuclear medium \cite{margueron04,ducoin06}. In this context, one would like to know whether the onset of the instabilities is strongly dependent on the properties of the underlying interaction. Astrophysical observations could then, in principle, provide constraints for nuclear effective interactions. Moreover, with over a hundred Skyrme force parameterizations at hand, it is nowadays possible to carry out ``statistical" calculations with a comprehensive set of different mean-fields \cite{chen09}. If one supposes that all these forces are equivalent, their predictions should span a physically relevant parameter space of observables within the mean-field approximation. Focusing on the liquid-gas phase transition, the predictions obtained for different interactions should provide a set of valid theoretical critical points. Similar studies, relating the liquid-vapor coexistence curve to the underlying interaction, have been performed in the context of chemical physics \cite{okumura00}. 

By computing the liquid-gas critical properties with large sets of mean-fields, one might also be able to identify behaviors which are independent of effective interactions. Examples of such correlations will be given in Section \ref{sec:sat}. In addition, a comprehensive study of the critical points predicted by forces with different saturation properties can assess in a systematic way the long-standing question of whether there is a connection between the critical and the saturation points. A link between these two physically relevant points might yield relevant information about the temperature dependence of the equation of state. Moreover, extending these results to the case of isospin asymmetric matter could provide a connection with recent studies on correlations between zero-temperature, isospin-asymmetric nuclear properties \cite{chen09}. Finally, in the context of critical exponents, one would expect these to be independent of the underlying mean-field. This can be tested numerically with self-consistent calculations of the liquid-gas phase transition, providing a nice example of how physics at the critical point is dominated by fluctuations. 

This paper is structured as follows. Section \ref{sec:HF} provides a brief summary of the formalism associated with the finite temperature self-consistent Hartree-Fock. The properties of the liquid-gas phase transition and the spinodal and coexistence curves are also introduced. In Section \ref{sec:models}, analytical models that provide a useful qualitative understanding for the liquid-gas phase transition are discussed. The relation between the saturation, flashing and critical points of nuclear matter is explored in Section \ref{sec:sat}. Section \ref{sec:critical} is devoted to analyze the law of corresponding states and the critical exponents of the phase transition for self-consistent mean-field calculations. The conclusions and outlook for future work are outlined in Section \ref{sec:conc}.

\section{Self-consistent mean-field calculations of hot nuclear matter}
\label{sec:HF}

The self-consistent Hartree-Fock method for nuclear matter calculations offers a theoretically consistent model for the description of hot nuclear matter \cite{vautherin96}. Physically, this formalism stems out from a minimization of the free energy of the system subject to the condition that the density matrix is of the one-body type \cite{bonche84}. Two- and higher-body correlations are therefore neglected on the basis that their contributions are captured by the parameterization of the effective interaction. Alternatively, one can cast the problem in terms of density functional theory, for which an energy density functional is guessed (based on some knowledge of the system) and a corresponding local density is found via a minimization of the energy (or, at finite temperature, the grand-potential \cite{mermin65,gupta80}).

In the following, results are presented for the nuclear matter liquid-gas phase transition with two different types of widely used effective interactions. The Skyrme force is a density-dependent, zero-range parameterization:
\begin{align}
V_{NN}({\overrightarrow r}) &= 
t_0 (1 + x_0 P_s) \delta({\overrightarrow r}) +
\frac {1}{6} t_3 (1 + x_3 P_{s}) \rho^{\alpha} \delta({\overrightarrow r})  \nonumber \\
&+ \frac {1}{2} t_1 ( 1 + x_1 P_s) 
\left[ { \overleftarrow k}^2 \delta({\overrightarrow r}) + \delta({\overrightarrow r}){\overrightarrow k}^2 \right]
\nonumber \\
&+ t_2 ( 1 + x_2 P_s) \left[ {\overrightarrow k} \cdot \delta({\overrightarrow r}) {\overrightarrow k} \right] \, , 
\label{eq:skyrme}
\end{align}
where ${\overrightarrow r}$ is the relative distance between two interacting nucleons and ${\overrightarrow k}= ({\overrightarrow \nabla}_1 -{\overrightarrow \nabla}_2)/2 i $ is their relative momentum acting on the right. ${\overleftarrow k}$ is its conjugate acting on the left, while $P_s= (1 + \overrightarrow{\sigma}_1 \cdot \overrightarrow{\sigma}_2)/2$ represents the spin exchange operator and $\rho$ corresponds to the total nucleonic density \cite{vautherin72}. The spin-orbit term is disregarded, since it does not contribute in homogeneous systems. Over a hundred parameterizations of this force exist in the literature \cite{stone03}. Results for 39 forces will be presented. 27 of these passed a series of criteria imposed in Ref.~\cite{stone03}: they offer a good description of both nuclear and neutron matter. In addition, to provide checks with previously published studies, the SIII, SkM$^*$, RATP and SkS3 parameterizations  \cite{chabanat97}, together with 2 forces of Ref.~\cite{song91} (forces SI and SJ1) and Ref.\cite{jaqaman83} (forces ZR1 and ZR2), have also been considered. Finally, more recent mean-field models, like the LNS parameterization \cite{lombardo06}, fitted to microscopic Brueckner--Hartree--Fock calculations, as well as the latest force of Goriely \emph{et al.}~\cite{goriely09}, which provides an excellent description of finite nuclei properties, have been included in the discussion.

The Gogny interaction has also been used extensively for mean-field and pairing calculations of nuclei and nuclear matter. In addition to a density dependent zero-range term, it includes a set of finite-range terms:
\begin{align}
V_{NN}({\overrightarrow r})&=
\sum_{i=1}^{2} e^{-\frac{r^2}{\mu_i^2}}
\Big( W_i+B_i\,P_s-H_i\,P_t-M_i\,P_s \, P_t \Big) \nonumber \\
&+ t_0 (1+x_0\,P_s) \rho^{\alpha} \delta ({\overrightarrow r}) \, .
\label{eq:gogny}
\end{align}
as well as a more complex spin-isospin structure (as seen by the additional isospin exchange operator, $P_t$). There are far fewer Gogny forces than Skyrmes in the literature. The following parameterizations will be considered: D1 \cite{gogny80}, D1S \cite{berger91}, D1P (with its additional zero-range term) \cite{farine99} and the very recent D1M force \cite{goriely09b}. The D250--D300 parameterizations of Ref.~\cite{blaizot95} will also be included in this analysis, since they might be useful in highlighting a relation between the compressibility and the critical properties of Gogny forces.

In contrast to the zero temperature case, for which the equations for the microscopic and macroscopic properties of nuclear matter can be cast analytically for both effective interactions, the self-consistent Hartree-Fock approximation needs to be implemented numerically at finite temperature. The formalism presented in the following is grand-canonical and the external fixed variables should be the temperature, $T$, and the chemical potential, $\mu$. In the thermodynamic limit, however, ensemble equivalence is guaranteed and one can work equivalently in a canonical picture, where the average density of the system, $\rho$, instead of $\mu$, is considered as an external variable. The energy density in the Hartree-Fock approximation is given by the sum of the kinetic term plus a potential term, obtained from the expectation value of the effective interaction between two plane waves of momentum $k_1$ and $k_2$:
\begin{align}
\frac{E}{\Omega} \left(\rho,T \right) &= \nu \int \frac{\textrm{d}^3 k}{(2 \pi)^3} f(k) \frac{\hbar^2 k^2}{2 m} \nonumber \\
&+ \frac{\nu^2}{2} \int \frac{\textrm{d}^3 k_1}{(2 \pi)^3} \frac{\textrm{d}^3 k_2}{(2 \pi)^3} 
 f(k_1) f(k_2) \langle \vec{k_1} \vec{k_2} | V_{NN} | \vec{k_1} \vec{k_2} \rangle_a \, ,
\label{eq:ener} 
\end{align}
where $\nu=4$ is the spin-isospin degeneracy of the system and the properly anti-symmetrized matrix elements have been averaged over spin and isospin. The momentum distribution, $f(k)$, describes the thermal population of momentum states according to a Fermi-Dirac distribution,
\begin{align}
f(k) &= \left[ 1+ e^{\left( \frac{\hbar ^2 k^2}{2m} + U(k) - \mu \right)/T} \right]^{-1} \, ,
\label{eq:FD} 
\end{align}
and depends on both the chemical potential, $\mu$, and the mean-field, $U(k)$. The chemical potential can be found by inverting the relation:
\begin{align}
\rho = \nu \int \frac{\textrm{d}^3 k}{(2 \pi)^3} f(k) \, ,
\label{eq:norm} 
\end{align}
which yields the normalization of the distribution to the total density of the system. The mean-field is given by:
\begin{align}
U(k) = \nu \int \frac{\textrm{d}^3 k'}{(2 \pi)^3} f(k') \langle \vec{k} \vec{k}' | V_{NN} | \vec{k} \vec{k}' \rangle_a \, ,
\label{eq:meanfield} 
\end{align}
and also depends on the momentum distribution, $f(k)$. As a result, Eqs.~(\ref{eq:norm}) and (\ref{eq:meanfield}) are coupled and need to be solved simultaneously, which gives rise to the self-consistent Hartree-Fock solution. Once the momentum distribution is set, it can be used to compute bulk thermodynamical properties, such as the entropy density:
\begin{align} 
\frac{S}{\Omega} (\rho,T)&= \nu \int \frac{d^3k}{(2\pi)^3} \left[ f(k) \ln f(k) + (1- f(k)) \ln(1- f(k)) \right] \, .
\label{eq:entro}
\end{align} 
From this and the energy of Eq.~(\ref{eq:ener}), the free energy is obtained, $F=E - TS$, and access to the pressure is gained by taking a derivative of the free energy per particle, $p=\rho^2 \frac{\partial F/A}{\partial \rho}$. To avoid the latter differentiation, one usually replaces the derivative with the chemical potential, so that $p= \rho(\mu - \frac{F}{A})$. This replacement relies on the thermodynamical consistency of the theory, which is only achieved if the rearrangement terms of the mean-field, originated by its density dependence, are properly considered in the inversion of Eq.~(\ref{eq:norm}) \cite{vautherin96}.

The solution of the Hartree-Fock equations is particularly easy for the Skyrme mean-field, due to its simple structure. Most importantly, the mean-field is independent of temperature and only depends quadratically on momentum. This allows for a separation of the type $U(k)=\bar U + \frac{\hbar^2 k^2}{2m} \Delta \rho $. The first term, including the rearrangement contribution, amounts to:
\begin{align}
\bar U  = \frac{3}{4} t_0 \rho + \frac{1}{16} ( \alpha + 2 ) t_3 \rho^{\alpha+1}\, ,
\end{align}
while the momentum dependent term can be easily incorporated in the definition of the effective mass:
\begin{equation}
\frac{m^*(\rho)}{m}= \frac{1}{ 1 + \frac{2m}{\hbar^2} \frac{1}{16} \left( 3 t_1 + 5 t_2 + 4 t_2 x_2 \right) \rho  } =\frac{1}{ 1 + \Delta \rho  } \, .
\label{eq:effmass}
\end{equation}
Note that, for Skyrme forces, the mean-field is independent of $f(k)$, $\bar U$ can be absorbed into the chemical potential and the effective mass contribution can be included in the kinetic term. As a consequence, the numerical solution of Eq.~(\ref{eq:norm}) is substantially simplified. In particular, even though at the diagrammatic level the calculation is self-consistent, in the numerical implementation there is no need for a self-consistency loop. 

The determination of the mean-field is more difficult with the Gogny force. In this case, due to the finite range of the interaction, the single-particle potential is momentum, density and temperature dependent:
\begin{align}
\bar U (k,\rho,T) &= \frac{\pi^{3/2} \rho}{4} \sum_{i=1}^2 \mu_i^3 \left[ 4 W_i + 2 B_i - 2 H_i - M_i\right] \nonumber \\
           	& - \pi^{3/2} \sum_{i=1}^2  \left[ W_i + 2 B_i - 2 H_i - 4 M_i\right] g_i( k, \rho, T )\nonumber \\
                &+ \frac{3}{8} t_0 (  \alpha + 2 ) \rho^{\alpha+1} \, .
\end{align}
The exchange integral, 
\begin{align}
g_i(k,\rho,T) &= \nu \int \frac{\textrm{d}^3 k'}{(2 \pi)^3} e^{- \frac{\mu_i}{4} |\vec k - \vec k'| } f(k') \, ,
\label{eq:exchange}
\end{align}
is responsible for the non-trivial temperature and momentum dependences. The Gogny mean-field cannot be separated cleanly into an effective mass and a momentum independent term. This separation might eventually be performed as an additional approximation and, on the whole, should provide reasonable results \cite{lopez06}. However, full mean-field calculations are preferable for accurate calculations of liquid-gas coexistence curves. Because of the implicit dependence of the exchange integral on the mean-field, a self-consistency loop at the computational level, absent in Skyrme calculations, is needed. 

\begin{figure}[t]
	\begin{center}
	\includegraphics[width=0.75\linewidth]{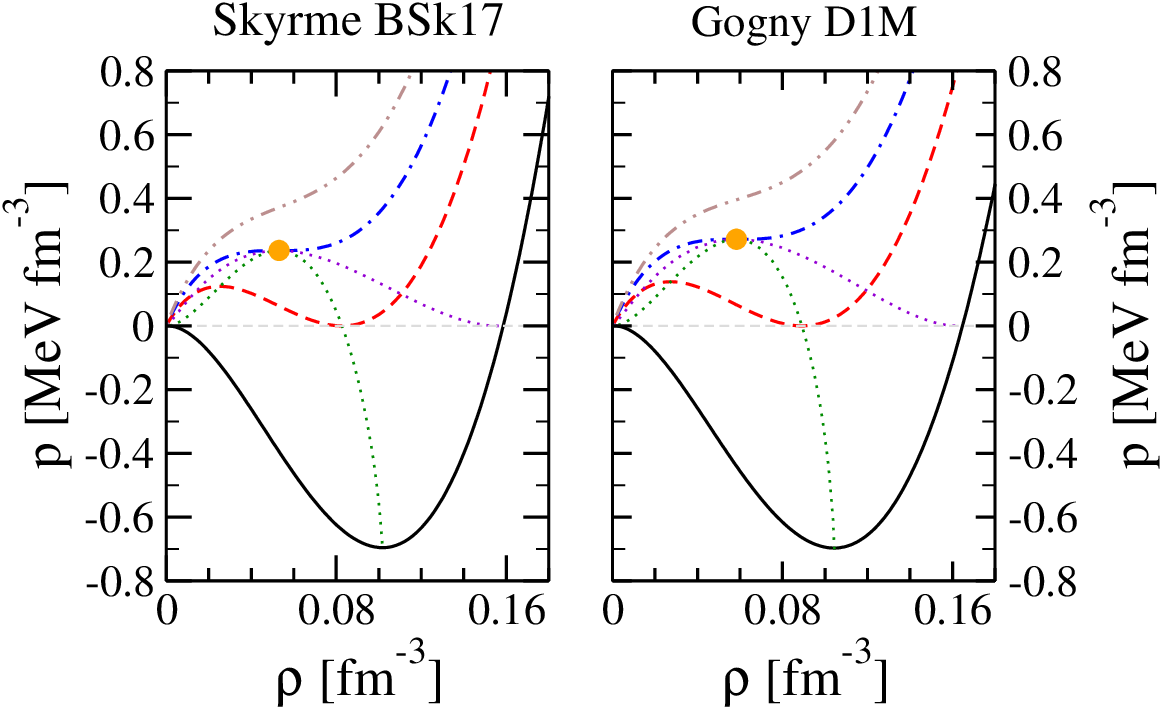}
	\caption{Pressure isotherms for different temperatures: $T=0$ (solid line), $T=T_f$ (dashed line), $T=T_c$ (dash-dotted line) and $T=18$ MeV (dash-double-dotted line).  The dotted lines represent the corresponding spinodal and coexistence regions. The left panel has been obtained with the BSk17 interaction, while the right panel shows results for the Gogny D1M force.}
	\label{fig:isoterms}
	\end{center}
\end{figure}

The density and temperature dependence of the thermodynamical properties obtained within the Hartree-Fock approximation are non-trivial. In particular, they cannot be parametrized in analytical form except for extreme regimes of degeneracy. The mean-field solution describes \emph{simultaneously} two different physical phases: a gas at low densities and a liquid at densities close to saturation \cite{bonche85}. This behavior is more easily visualized in the pressure isotherms of nuclear matter, as show in Fig.~\ref{fig:isoterms}. Left and right panels correspond to results obtained with typical Skyrme (BSk17) and Gogny (D1M) forces, respectively. Four characteristic temperatures have been chosen for illustrative purposes. 

At $T=0$ (solid lines), the pressure is a decreasing function of the density at low densities until a minimum at $\sim \rho_0/2$ (where $\rho_0$ is the saturation density) is reached. This regime, for which $\frac{\partial p}{\partial \rho} < 0$, corresponds to the spinodal region, where the system is mechanically unstable. As temperature increases, the spinodal region, which is shown in light dotted lines in Fig.~\ref{fig:isoterms}, shrinks. At low densities, a mechanically stable gas phase forms and coexists with the higher density liquid phase. The spinodal region is associated with negative pressures at low temperatures. This picture changes as temperature increases and the thermal contribution tends to push the pressure to largest values. The isotherm at which the pressure becomes a completely positive function defines the so-called \emph{flashing} temperature. More specifically, the flashing point satisfies the two simultaneous constraints:
\begin{align}
p = \frac{\partial p}{\partial \rho} = 0 \, .
\label{eq:flashing}
\end{align}
(Note that, for nuclear matter, the point at $T=0$ and $\rho=0$ also satisfies this condition). In Fig.~\ref{fig:isoterms}, the flashing point corresponds to the crossing of the spinodal with the $p=0$ line. For an isolated system, without an external gas to stabilize it, this would correspond to the maximum temperature at which the system could still be self-bound The definition of the flashing temperature, $T_f$, is therefore somewhat similar to that of the limiting temperature of a nucleus, \emph{i.e.}~the maximum temperature a nucleus can withstand before evaporating into a gas of nucleons. As a matter of fact, $T_f$ might be considered as the extrapolation of the limiting temperature to the $A \to \infty$ case.

For the infinite system, however, the system can exist above the flashing temperature because the gas and the liquid phase exert the same pressure on each other. According to the Maxwell criterion, the condition that the gas and liquid phases have the same pressure, together with the requirement that their chemical potentials are the same, set the coexistence points \cite{huang,pathria}. The coexistence region is shown with a dark dotted line in Fig.~\ref{fig:isoterms}. Note that, at $T=0$, there is a coexistence between a zero-density gas and a liquid at saturation density. Increasing the temperature results into larger gas and smaller liquid densities. Eventually, the two densities meet at the critical point, which occurs at the critical temperature, $T_c$. The critical isotherm is shown with a dashed-dotted line in both panels of Fig.~\ref{fig:isoterms}. The critical point is found at the endpoint of both the spinodal and the coexistence lines, and therefore can be found by requiring:
\begin{align}
\frac{\partial p}{\partial \rho} =  \frac{\partial^2 p}{\partial \rho^2} = 0 \, .
\label{eq:critical}
\end{align}
Above the critical temperature, the system will only exist in the gas phase and the pressure isotherms become a monotonically increasing function of density. An example of this behavior is given by the dashed-double-dotted lines, which have been computed at $T=18$ MeV for the two panels.

For optimal liquid-gas calculations, it is important to compute Fermi integrals [integrals over the Fermi-Dirac distributions such as those in Eqs.~(\ref{eq:ener}), (\ref{eq:norm}), (\ref{eq:meanfield}) and (\ref{eq:entro})], with great accuracy and, simultaneously, short computational times. Numerical techniques based on Ref.~\cite{aparicio98} have been implemented to this end. For consistency, the critical point has been determined by three different numerical methods. The first two estimates have been obtained by finding the endpoint of the coexistence and the spinodal curves, respectively. These curves are computed separately using Newton-Raphson routines and extrapolations are performed to determine their maximum accurately. In addition, the last temperature and density where the first and second derivatives of the pressure show a node has been found using a bisection method. The three different methods coincide in the determination of the critical temperature to the second decimal point for all the mean-fields considered here. For the forces where it was possible to do so, the results presented have been checked against those previously published in the literature. To my knowledge, this is the first time that a consistent and comprehensive set of mean-fields is used to find correlations between physical parameters in the liquid-gas phase transition. 

\section{Models for the phase transition}
\label{sec:models}

\begin{figure}[t]
	\begin{center}
	\includegraphics[width=0.75\linewidth]{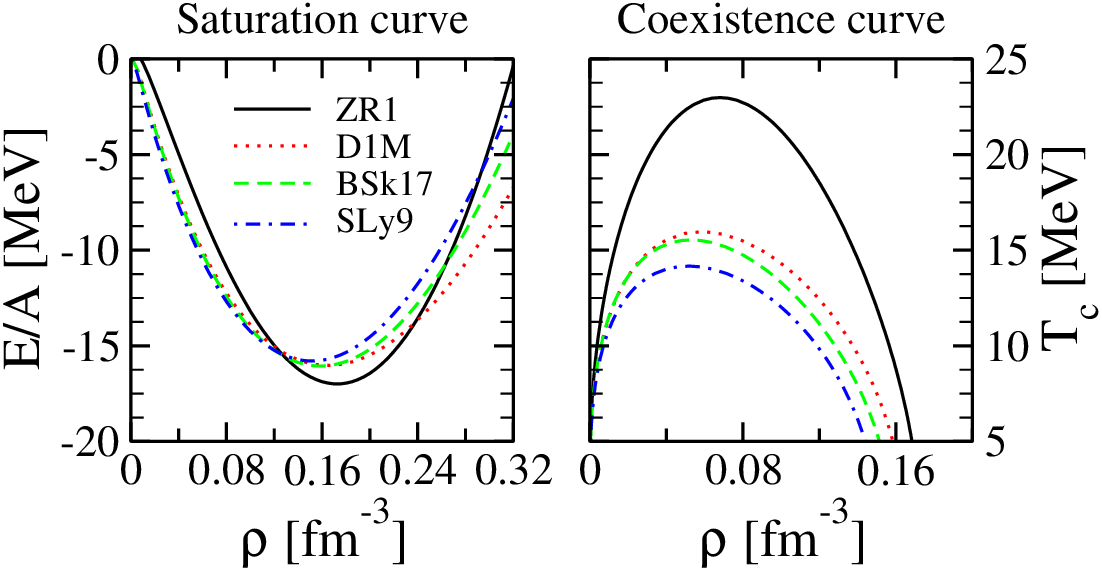} 
	\caption{Left panel: saturation curve of nuclear matter for four different interactions. Right panel: liquid-gas coexistence curves for the same interactions.}
	\label{fig:sat_coex}
	\end{center}
\end{figure}

Finite temperature nuclear studies provide new means of understanding the properties of the dense nuclear medium. With the suitable theoretical framework, the information gathered from hot nuclei could be used to constraint the nuclear mean-field. In particular, two points of the phase diagram of nuclear matter stand out for their relevance. On the one hand, the saturation point defines the properties of cold nuclear matter and is directly related to nuclear structure data. On the other hand, the liquid-gas critical point can only be accessed if excited matter is produced in a reaction. Can one find a direct connection between the critical and saturation points? In principle, the partition function of an interacting system is not constrained by its zero temperature properties. For nuclear matter within the mean-field picture, however, the effective interaction determines the energy and also the entropy [via the single-particle potentials in Eq.~(\ref{eq:entro})] and perhaps this indirect connection might be strong enough to provide a correlation between such properties.

At first sight, however, such a connection is not straightforward. An example is provided by Fig.~\ref{fig:sat_coex}, where the saturation curve (left panel) and the liquid-gas coexistence curve (right panel) are shown for four different mean-fields. The Skyrme forces ZR1 and SLy9 have the highest and lowest critical temperatures, respectively, of the interactions considered here. D1M and BSk17 should be taken as representatives of the ``average" behavior of Gogny and Skyrme forces. Even though these four interactions have relatively similar saturation points, with saturation energies deviating by less than $1$ MeV from one another, the corresponding critical temperatures span a wide range of values, from $14$ to $23$ MeV. This suggests that no connection exists between the saturation and the critical point.

In contrast to this naive analysis, data concerning nuclear caloric curves has been used to provide constraints on the properties of nuclear matter at the saturation point \cite{natowitz02b}. The knowledge gathered in this direction, however, has been generally based on models that connect saturation and critical properties by assuming analytical density and temperature dependences of the equation of state. By construction, the thermodynamical properties within these models have underlying temperature and density dependences that might or might not be good approximations of the full nuclear system. Generally, such models are based on expansions in particular degeneracy regimes. The degeneracy parameter, $d=\frac{T}{\varepsilon_F}$, where $T$ is the temperature and $\varepsilon_F$, the Fermi energy, measures the relative importance of thermal effects against quantum fluctuations. The thermodynamical properties in the Hartree-Fock approximation can be found analytically at the extremes of degeneracy. On the one hand, in the degenerate regime ($d << 1$), \emph{i.e.}~at relatively low temperatures or high densities, the Sommerfeld expansion provides a systematic expansion of the thermodynamical properties in terms of $d$ \cite{pathria}. This yields the well-known quadratic (linear) temperature dependence for the energy (entropy). On the other hand, in the semiclassical limit ($d >> 1$), \emph{i.e.}~for very dilute and hot systems, the fugacity expansion provides a connection with classical statistical mechanics \cite{huang}. 

Notions of both the degenerate and the semiclassical limits have been used in the literature to model the behavior of the hot equation of state \cite{jaqaman83,kapusta84}. Unfortunately, the actual nuclear liquid-gas phase transition takes place at $T \sim 15$ MeV and $\rho \sim 0.08$ fm$^{-3}$, so that $d \sim \frac{1}{3}$ and none of the above expansions can be applied with extreme confidence. Nowadays, full self-consistent calculations can be implemented numerically and very accurate results are found in extremely short computational times (seconds for the case of Skyrme forces). Consequently, there is no need to rely on such expansions other than to get a physical insight on the properties of liquid-gas transition. In the following, I shall discuss some of these models and analyze their validity for the liquid-gas phase transition. In particular, I will look at the relation between saturation and critical parameters. Incidentally, these models will also provide insight into a direct connection between the flashing and the critical points, as will be discussed in Section \ref{sec:sat}.

\subsection{Jaqaman model}

A very simple model that already encodes some of the physically relevant information associated with the nuclear liquid-gas transition is provided by the lowest order fugacity expansion with $m^*=m$. This model (and higher order contributions in the expansion) were discussed by Jaqaman \emph{et al.} in Ref.~\cite{jaqaman83}. Within this approach, the temperature dependence of the pressure is taken to be that of a classical gas and the density dependence is supplemented with the zero-range terms of a Skyrme force:
\begin{align}
p = \rho T + \frac{3}{8} t_0 \rho^2 + \frac{1}{16} t_3 (\alpha+1) \rho^{\alpha+2} \, .
\label{eq:jaqaman}
\end{align} 
This expression neglects any quantum fluctuation (in the sense that the temperature dependence is purely classical) and, strictly speaking, should only be valid in the semiclassical regime. The flashing point can be obtained analytically from Eq.~(\ref{eq:flashing}):
\begin{align}
\rho_f &= \left[ \frac{6}{(\alpha+1)^2} \frac{t_0}{t_3} \right]^{1/\alpha} \, , \label{eq:jaqaman_rf} \\
T_f &= \frac{3}{8} \frac{\alpha}{\alpha+1} t_0 \rho_f \, ,
\label{eq:jaqaman_tf}
\end{align}
while the critical point, deduced from Eq.~(\ref{eq:critical}), reduces to:
\begin{align}
\rho_c &= \left[ \frac{12}{(\alpha+1)^2 (\alpha+2)} \frac{t_0}{t_3} \right]^{1/\alpha} \, , \label{eq:jaqaman_rc} \\
T_c &= \frac{3}{4} \frac{\alpha}{\alpha+1} t_0 \rho_c \, , \label{eq:jaqaman_tc} \\
p_c &= \frac{\alpha+1}{2(\alpha+2)} \rho_c T_c \, . \label{eq:jaqaman_pc}
\end{align}
This expression gives an important insight into the relation between the critical point and the underlying equation of state. The critical density is determined by the ratio of $t_0$ and $t_3$, so it will be sensitive to the interplay of the attractive and the repulsive components of the force. Moreover, $\rho_c$ depends non-analytically on $\alpha$. The latter point stresses the importance of the density dependence of the force in determining the critical point. In particular, the compressibility, $K_0$, is to a great extent fixed by $\alpha$ and, thus, it will play a role in the determination of the critical point \cite{natowitz02b,blaizot95}. The so-called critical compressibility factor is given by the dimensionless parameter $\gamma_c = \frac{p_c}{T_c \rho_c}$ and it is useful in characterizing the critical point \cite{huang}. For the van der Waals equation of state, $\gamma_c=\frac{3}{8}=0.375$ while, for the Jaqaman model, $\gamma_c=\frac{\alpha+1}{2(\alpha+2)}$. For Skyrme forces, which usually have $\alpha=0 - 1$, this corresponds to lower critical compressibility factors than the van der Waals prediction, between $\frac{1}{4} \leq \gamma_c \leq \frac{1}{3}$. 

Within the Jaqaman model, the critical and the flashing points are closely connected. Their ratios are purely determined by the exponent of the density dependence:
\begin{align}
\frac{\rho_c}{\rho_f} &= \left[ \frac{2}{\alpha+2} \right]^{1/\alpha} \, , \label{eq:jaqaman_r} \\
\frac{T_c}{T_f} &= 2 \frac{\rho_c}{\rho_f} \, . \label{eq:jaqaman_t}
\end{align}
For Skyrme mean-fields, this results in a very limited range of allowed ratios: $\frac{1}{\sqrt{e}} \leq \frac{\rho_c}{\rho_f} \leq \frac{2}{3}$. Accordingly, the quotient between critical and flashing temperatures is twice as large, $\frac{T_c}{T_f} \sim 1.3$. 

The Jaqaman model provides a very poor description of the liquid branch. The latter involves densities close to saturation and therefore sits closer to the degenerate regime. Specifically, the linear temperature dependence of the pressure is only valid at extremely low densities and therefore it is too simple to explain accurately the whole region involved in the transition. This simplistic parameterization of the finite temperature properties leads to large differences between the critical temperature of the full self-consistent calculation and those of the respective Jaqaman models (see Fig.~\ref{fig:jaqaman}).

\subsection{Kapusta model}

The expansion near the completely degenerate case can also be used to establish the temperature dependence of the equation of state. The Kapusta model, originally derived in Ref.~\cite{kapusta84}, relies on the Sommerfeld expansion together with the knowledge of the compressibility to build an approximation for the pressure of the system:
\begin{align}
p = \frac{K_0}{9} \frac{\rho^2}{\rho_0^2} \left(\rho - \rho_0 \right) + \frac{m_0^*}{6} b^2 \rho^{1/3} T^2 \, .
\label{eq:kapusta_pres}
\end{align}
Ingredients of this model include the numerical constant $b=\left(\frac{2^{5/2} \pi}{3 \hbar^3} \right)^{1/3}$ and the effective mass at saturation density, $m_0^*=m^*(\rho_0)$. The density dependence of the latter is neglected, since $m_0^*$ is basically introduced to tune the thermal density of states close to saturation. Again, analytical formulae can be found for both the flashing:
\begin{align}
\rho_f &= \frac{5}{8} \rho_0 \, , \label{eq:kapusta_rf} \\
T_f &= \frac{5^{5/6}}{2^{7/2} b} \sqrt{\frac{K_0}{m_0^*} }  \rho_0^{1/3} \, ,
\label{eq:kapusta_tf}
\end{align}
and the critical points:
\begin{align}
\rho_c &= \frac{5}{12} \rho_0 \, , \label{eq:kapusta_rc} \\
T_c &= \frac{5^{5/6}}{2^{13/6} 3^{1/3} b} \sqrt{\frac{K_0}{m^*_0}} \rho_0^{1/3} \, , \label{eq:kapusta_tc}\\
p_c &= \frac{25}{324} K_0 \rho_c   \label{eq:kapusta_pc} \, .
\end{align}
Plugging in typical values of saturation parameters ($\rho_0=0.16$ fm$^{-3}$, $K_0 \sim 230$ MeV and $m^*_0=0.7 m$), the critical point is predicted to be at $\rho_c=0.066$ fm$^{-3}$, $T_c=20.7$ MeV and $p_c=1.18$  MeVfm$^{-3}$. While $\rho_c$ and $T_c$ are rather typical and comparable to mean-field values (see Table \ref{table} below), $p_c$ is somewhat large. This suggests that the pressure isotherms in this model overestimate the mean-field ones. As a result, the critical compressibility factor becomes extremely large, $\gamma_c = 0.89$, in contrast to the van der Waals and mean-field values.  

Within the Kapusta model, both $\rho_f$ and $\rho_c$ are directly proportional to the saturation density. This suggests that a clean connection between critical and saturation densities might be established. Mean-field calculations do not support these findings, as will be discussed later on (see Fig.~\ref{fig:sat_coex}). Just as in the previous model, however, the critical and flashing points are closely related to each other:
\begin{align}
\frac{\rho_c}{\rho_f} &= \frac{2}{3} \, , \label{eq:kapusta_r} \\
\frac{T_c}{T_f} &= \left(\frac{16}{3} \right)^{1/3} \, . \label{eq:kapusta_t}
\end{align}
The ratio of densities coincides with the upper value predicted by the Jaqaman model.  The flashing and critical temperatures are functions of the combination $\eta=\sqrt{\frac{K_0}{m_0^*}} \rho_0^{1/3}$, so that their ratio is a simple numerical factor. As a matter of fact, $\frac{T_c}{T_f} =1.75$, which is substantially larger than the value predicted by the Jaqaman model. 

The dependence of $T_f$ and $T_c$ on $\eta$ is helpful in understanding the links between the saturation and critical points. To begin with, stiffer equations of state lead to larger critical temperatures. At $T=0$, the pressure in the Kapusta model has a minimum value of $p_{\textrm{min}}= - \frac{4}{81} K_0 \rho_0$ at sub-saturation densities. High values of the compressibility or larger saturation densities are therefore associated with more negative pressures below saturation. Consequently, higher temperatures are needed before the pressure can become zero (flashing point) or its minimum disappears (critical point). This explains why $T_c$ and $T_f$ are monotonically increasing functions of $K_0$ and $\rho_0$. Similarly, ``stronger" temperature dependences (in the sense that the density of states is larger due to an increase in $m^*_0$) can overcome negative pressures easily. Larger effective masses are therefore associated with lower critical and flashing temperatures. 

The Kapusta model is based on a degeneracy expansion which is only valid at low temperatures and large densities. As a result, the pressure isotherms of Eq.~(\ref{eq:kapusta_pres}) are very close to the mean-field ones at $T=0$. As temperature is increased and the critical point is approached, however, the agreement worsens. The origin of these discrepancies can be traced back to the simplified temperature dependence, but also to the lack of density dependence in the density of states. The former effect can be improved by considering higher order terms in the degeneracy expansion, while the latter is likely to play an important role for the liquid-gas phase transition, since the flashing and critical points are largely determined by the density derivatives of $p$ [see Eqs.~(\ref{eq:flashing}) and (\ref{eq:critical})]. An extension of the Kapusta model which includes a more realistic effective mass dependence is discussed in Section \ref{sec:extendedkapusta}. 

\subsection{Lattimer-Swesty and Natowitz predictions}

Two popular parameterizations of the critical temperature have been inspired by the Kapusta model relation between $T_c$ and the saturation properties. Lattimer and Swesty introduced in Ref.~\cite{lattimer91} a formula for the critical temperature in terms of saturation properties:
\begin{align}
T_c &= C_{LS} \sqrt{K_0} \rho_0^{-1/3} \, . 
\label{eq:lattimer_t}
\end{align}
While based on the Kapusta model, this expression has the opposite dependence on saturation density \emph{i.e.}~larger saturation densities lead to lower critical temperatures. The value for the numerical parameter, $C_{LS}=0.608$ MeV$^{1/2}$fm$^{-1}$, is obtained by fitting the critical temperature of a series of simple mean-field models \cite{lattimer91}. 

Later on, Natowitz \emph{et al.} generalized the previous expression to include the effect of the effective mass near saturation, $m_0^*$:
\begin{align}
T_c &= C_N \sqrt{ \frac{K_0}{m_0^*/m} } \rho_0^{-1/3} \, . 
\label{eq:natowitz_t}
\end{align}
The empirical value of the parameter $C_N=0.484 \pm 0.074$ MeV$^{1/2}$fm$^{-1}$ is determined from an average of several theoretical values \cite{natowitz02b}. In contrast to the Jaqaman or the Kapusta models, which try to describe the thermodynamical potential, these two predictions are phenomenological estimates that provide a fit to a series of critical point data. The extent to which these predictions is satisfactory will therefore depend on the original data used in the fits. In the following Section, the accuracy with which these formulae reproduce the nuclear matter mean-field results with Skyrme and Gogny forces will be discussed. 

\subsection{Extended Kapusta model}
\label{sec:extendedkapusta}
The last model to be presented here is an extension of the Kapusta model that, to my knowledge, has never been considered. It provides a particularly good prediction of the critical temperature in terms of saturation properties for mean-field calculations. A common feature of the models presented so far is the fact that they completely neglect the density dependence of the effective mass. The latter changes from values of around $\frac{m_0^*}{m}=0.7$ at saturation to $1$ at $\rho=0$. Since the effective mass sets the density of states at each density, this (up to) $30$ \% effect might be particularly important for the density and temperature dependence of the equation of state and, consequently, for accurate predictions of critical parameters. Moreover, the effective mass (particularly its energy dependence) has been acknowledged as one of the key parameters in determining nuclear caloric curves \cite{sobotka04,shetty09}.

To include this effect schematically, one can use as a guiding point the density dependence of $m^*(\rho)$ coming from the Skyrme interaction, Eq.~(\ref{eq:effmass}). This dependence enters the Kapusta model in the thermal term and modifies the expression of the pressure:
\begin{align}
p = \frac{K_0}{9} \frac{\rho^2}{\rho_0^2} \left(\rho - \rho_0 \right) + \frac{m}{6} b^2 
\frac{1+\frac{5}{2} \Delta \rho}{(1+ \Delta \rho)^2} \rho^{1/3} T^2 \, .
\label{eq:exkap_pres}
\end{align}
Note that this differs explicitly from Eq.~(\ref{eq:kapusta_pres}) even at $\rho=\rho_0$. In the original derivation of the Kapusta model, the only density dependence in the $T^2$ term comes from the Fermi energy in the denominator. The extended model includes more complicated dependences that arise from the density derivatives associated with the calculation of the pressure. 

Given this expression, one should go ahead and keep the density dependence in the solution of the flashing and the critical point equations. Unfortunately, the latter task is complicated by the fact that analytical solutions cannot be found. However, since both the flashing and critical points occur well below saturation density, the effective mass is close to the free mass and the dimensionless parameter $\Delta \rho$ is of order $10$ \% or so. One can then expand all the expressions in terms of this parameter and solve the corresponding equations to an accuracy of $\mathcal{O}( \Delta^2 \rho^2)$. Using this method, the flashing point is found to be:
\begin{align}
\rho_f &= \widetilde{\rho_f} \left[ 1 - \frac{9}{80} \Delta \widetilde{\rho_f} \right] \, , \label{eq:exkapustarl} \\
T_f &= \widetilde{T_f} \left[ 1 - \Delta \widetilde{ \rho_f } \right] \, , \label{eq:exkapustatl}   
\end{align}
where the tilded parameters correspond to those of the original Kapusta model with $m^*_0=m$. Similarly, the critical point becomes:
\begin{align}
\rho_c &= \widetilde{\rho_c} \left[ 1 - \frac{9}{20} \Delta \widetilde{\rho_c} \right] \, , \label{eq:exkapustar} \\
T_c &= \widetilde{T_c} \left[ 1 - \Delta \widetilde{ \rho_c } \right] \, , \label{eq:exkapustat} \\
p_c &= \widetilde{p_c} \left[ 1 - \Delta \widetilde{ \rho_c } \right] \, . \label{eq:exkapustap}  
\end{align}

For most Skyrme parameterizations, the effective mass decreases with density, so that $\Delta > 0$. As a consequence, both the critical and flashing densities are reduced by a very small numerical factor. The effective mass density dependence is more important on the flashing and critical temperatures. Note that, interestingly enough, and up to first order, both temperatures can be written as:
\begin{align}
T_f &= \frac{m^*(\widetilde{\rho_f})}{m} \widetilde{T_f}  \, , \label{eq:exkapustaeffl}  \\
T_c &= \frac{m^*(\widetilde{\rho_c})}{m} \widetilde{T_c}  \, . \label{eq:exkapustaeff} 
\end{align}
These expressions suggest that the effective mass below saturation is important for a proper determination of the flashing and critical temperatures. Generally, for Skyrme forces both effective masses are around $\sim 0.85-0.9$, which leads to a sizable effect, particularly if one is looking at accurately predicting critical values from saturation properties. The Skyrme force results presented in the following have been obtained with Eq.~(\ref{eq:exkapustaeff}). Moreover, these expressions have the virtue of being easily generalizable to effective interactions that have other (presumably more complex) density dependences. Assuming that the temperature dependence of the effective mass can be neglected, as is generally the case for Gogny forces, one can compute the effective mass at the Fermi surface at $T=0$ and plug it into these expressions. The examples with the Gogny force discussed in the following have been obtained in this way. Finally, these formulae can also be used in calculations performed with sophisticated many-body theories.

\section{Relation between saturation and critical properties}
\label{sec:sat}
\begin{figure}[th]
	\begin{center}
	\includegraphics[width=0.7\linewidth]{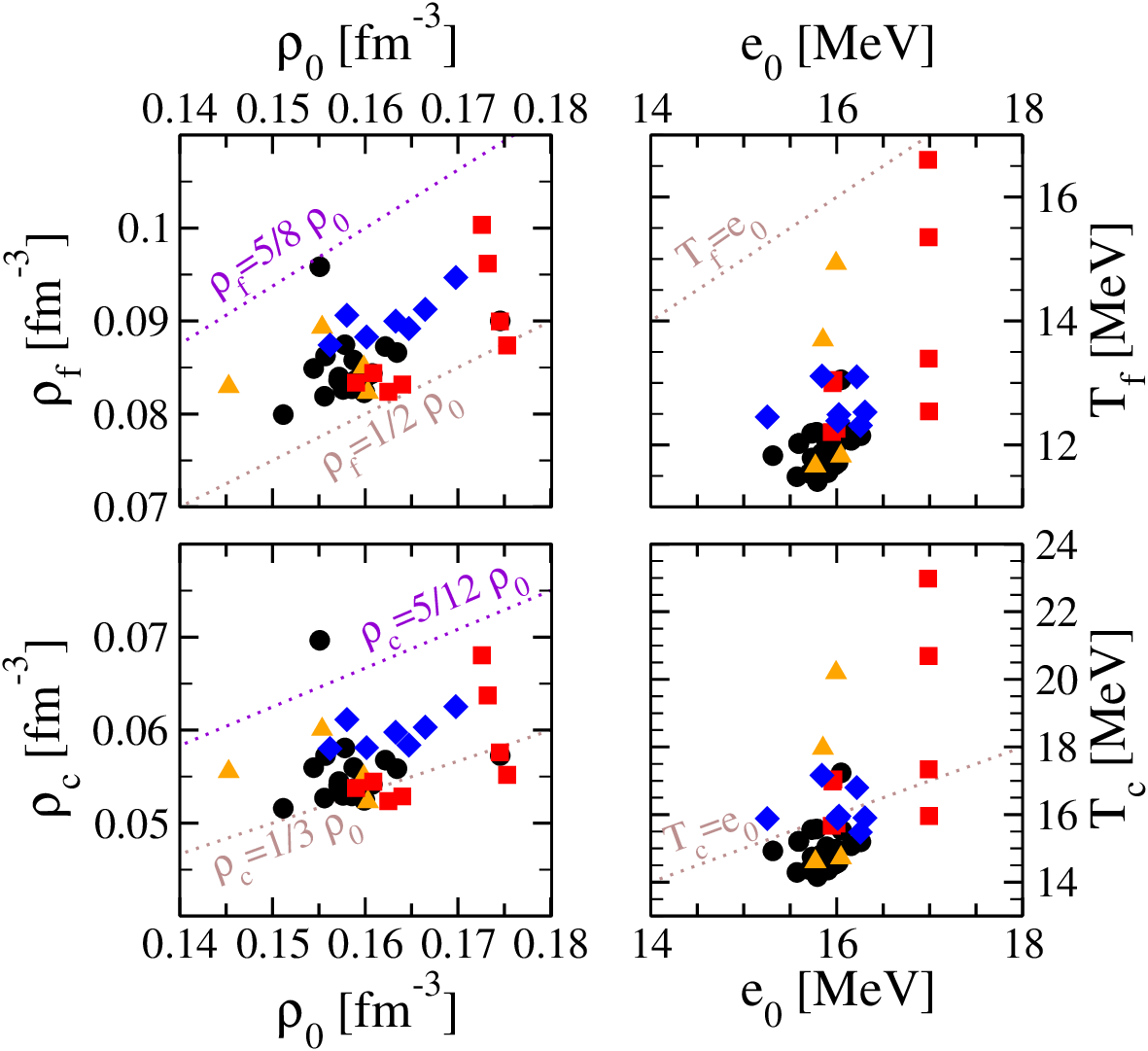}
	\caption{Left panels: flashing (upper panel) and critical (lower panel) densities versus saturation density for different effective interactions. Right panels: flashing (upper panel) and critical (lower panel) temperatures versus saturation energy. Skyrme interactions with $\frac{m^*}{m} = 1$ are represented with squares. The circles correspond to results obtained with modern Skyrme forces that have $\frac{m^*}{m} \neq 1$, while older forces are shown with triangles (see text for details). The diamonds have been obtained with the Gogny interaction.}
	\label{fig:corr_sat}
	\end{center}
\end{figure}

In this Section, the relation between the saturation and critical parameters predicted by self-consistent mean-field models will be analyzed by means of systematical calculations with an extensive set of effective interactions. Critical and flashing points will be deduced and compared to the predictions given by the models of the previous section. Thanks to the large number of mean-field parameterizations at hand, correlations between different parameters will be identified. With this, I will highlight the advantages and disadvantages of the different models and will try to draw a coherent picture of the liquid-gas transition properties as derived from accurate self-consistent calculations. In addition, this analysis is useful in assessing the correctness of naive dimensional arguments that have been put forward in the context of the liquid-gas phase transition. Note, however, that these correlations might not be physically realistic, since the description of the liquid-gas transition close to the critical point within a mean-field picture can be put into question \cite{Finocchiaro96}. 

The upper panels of Fig.~\ref{fig:corr_sat} concentrate on the relation between the flashing and the saturation points. The flashing points have been determined for several Skyrme (circles, triangles and squares) and Gogny (diamonds) forces. For reasons that will become clear later on, Skyrme interactions with no effective mass ($\frac{m^*}{m}=1$ or $\Delta=0$) are shown with squares. The circles correspond to modern Skyrme forces, \emph{i.e.} those with $m^* \neq m$ that passed the restrictions of Ref.~\cite{stone03} or those that have been introduced later on (LNS and BSk17). Triangles correspond to somewhat older forces (SI, SIII, SkM* and RATP). The upper left panel shows a scatter plot of the flashing densities obtained with finite temperature self-consistent calculations for different forces against their respective saturation densities. Any correlation between these parameters should appear as a clear trend in these figures. 

In most models, the saturation density is set (or fitted, depending on the particular model) to a value between $0.145$ and $0.175$ fm$^{-3}$. The range of allowed flashing densities is relatively narrower and goes from around $0.08$ to $0.10$ fm$^{-3}$. Actually, all $\rho_f$'s fall between two ``empirical" correlations lines. Limiting by below is the relation $\rho_f = 0.5 \rho_0$, whereas the upper limit is set by the Kapusta model prediction, $\rho_f = 0.625 \rho_0$. At first sight, however, it is hard to find a clear proportionality, as suggested by such relations. As a matter of fact, forces with the same saturation densities lead to flashing densities that differ by more than $0.02$ fm$^{-3}$, suggesting that additional physical information is needed if both densities are to be connected.

A somewhat similar situation is found by comparing the flashing temperature and the saturation energy, $e_0$ (upper right panel). Naively, one could have argued that the temperature needed to vaporize completely a liquid might be proportional to the binding energy of the latter. In contrast to this, none of the models presented in the previous Section predict an explicit dependence on $e_0$. On top of that, no clear proportionality trend between $T_f$ and $e_0$ is observed in Fig.~\ref{fig:corr_sat}. Actually, for all forces, the flashing temperatures are well below the corresponding saturation energies. Furthermore, for mean-fields that have very similar saturation energies, the corresponding flashing energies differ by more than $4$ MeV.

The lower panels of Fig.~\ref{fig:corr_sat} focus on the connection between the critical and the saturation densities. Similarly to what is observed for the flashing density, most of the $\rho_c$'s are found in a narrow range of values. A band, bound by below by the ``phenomenological" law, $\rho_c = \frac{1}{3} \rho_0$, and by above by the Kapusta model prediction, $\rho_c = \frac{5}{12} \rho_0 = 0.4166 \rho_0$, describes well most of the data. Yet, it is hard to extract a clear proportionality law from this picture. The lower correlation seems to describe particularly well some of the critical-saturation points, but one should not take the correlation $\rho_c = \frac{1}{3} \rho_0$ as a faithful prediction for critical densities. Similarly, looking at the lower right panel of Fig.~\ref{fig:corr_sat}, some critical temperatures happen to lie close to the $T_c=e_0$ line, but this law is hardly generalizable to all effective interactions. An even more striking fact is the noticeable spread in $T_c$: while all saturation energies lie within $2$ MeV from each other, critical temperatures span a range of almost $9$ MeV. This indicates that there is no connection between the saturation energy and the position of the critical point. It is interesting to note that the magnitude of the spread in $T_c$ is somewhat similar to the differences in critical temperatures found by using different realistic interactions and many-body techniques in Ref.~\cite{rios08}.

\begin{figure}[t]
	\begin{center}
	\includegraphics[width=0.75\linewidth]{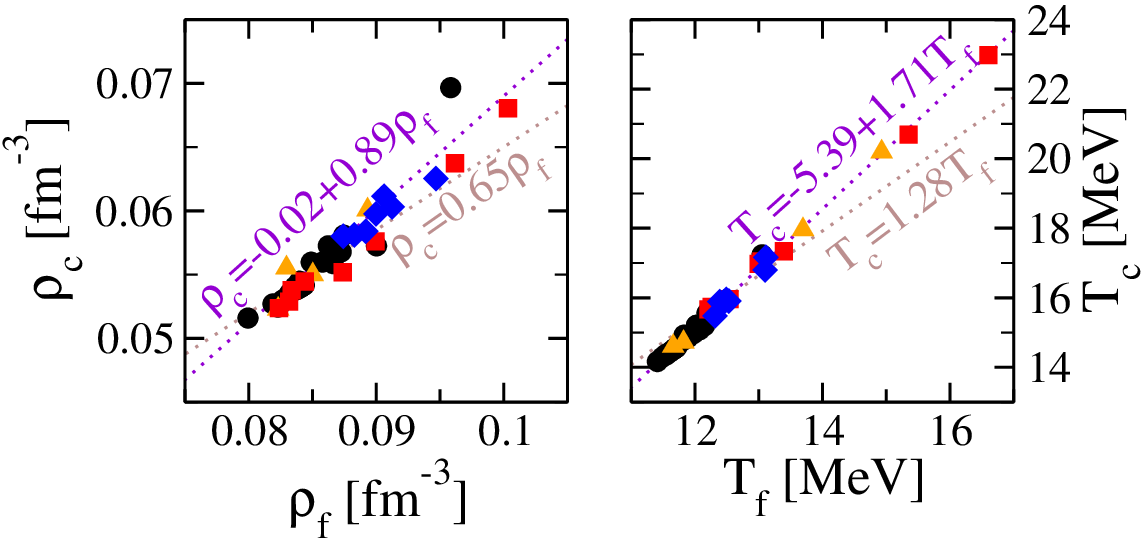}
	\caption{Left panel: critical density versus flashing density for different effective interactions. Right panel: critical temperature versus flashing temperature for different effective interactions.}
	\label{fig:crit_flash}
	\end{center}
\end{figure}

The scatter plots in the upper and lower panels of Fig.~\ref{fig:corr_sat} show very similar structures, with a clustering of Skyrme values around the central region, a more dilute set of Gogny forces and some $m^*=m$ forces in the right end. It appears that subsets of forces with analogous saturation properties behave similarly in their prediction for critical and flashing points. Accordingly, critical and flashing properties for different forces could be correlated. Such correlations are explored in the two panels of Fig.~\ref{fig:crit_flash}. In the left panel, critical densities obtained with different mean-fields have been plotted against their corresponding flashing densities. As suggested by both the Jaqaman and the Kapusta models [see Eqs.~(\ref{eq:jaqaman_r}) and (\ref{eq:kapusta_r})], a clear linear proportionality is found between the two. Fitting this linear relation with a single parameter, the law $\rho_c= 0.65 \rho_f$ is found, in good agreement with the proportionality constants predicted by the Kapusta and Jaqaman models. A more accurate fit of the larger densities can be achieved by considering an offset. The corresponding linear regression yields the relation $\rho_c= -0.02 + 0.89 \rho_f$. 

Similarly, a tight proportionality between the critical and the flashing temperatures is observed in the right panel of Fig.~\ref{fig:crit_flash}. A least-squares fit to a single slope parameter results in the relation $T_c= 1.28 T_f$, which describes well the low temperature points. Note that this constant of proportionality is close to the Jaqaman model prediction, Eq.~(\ref{eq:jaqaman_t}), and suggests that the relation $\frac{\rho_c}{\rho_f} = 2 \frac{T_c}{T_f}$ is fulfilled. The somewhat poor description of the high temperature points can be substantially improved by allowing for an offset. The linear relation $T_c= -5.39 + 1.71 T_f$ gives an excellent description of all the data. In this case, the slope is closer to that predicted by the original Kapusta model. Note, however, that the 5 highest critical temperatures correspond to forces with $m^*=m$ or with a poor description of asymmetric matter according to Ref.~\cite{stone03}. Modern Skyrme and Gogny forces seem to favor the regions $T_c \sim 14-17$ MeV and $T_f \sim 11-13$ MeV.

The tight linear correlation between $T_c$ and $T_f$ provides a clean connection between two seemingly unconnected points of the liquid-gas transition phase diagram. The correlation is generic, in the sense that it is valid for all the mean-fields considered. These findings exemplify the potential of using comprehensive sets of interactions to assess the physically allowed parameter spaces of observables. The correspondence between $T_f$ and $T_c$ might be useful in relating flashing and critical temperatures in other many-body approaches, particularly if one of the two temperatures cannot be accessed \cite{baldo04,rios08}. Note that theoretical correlations between limiting temperatures of nuclei and critical temperatures in nuclear matter have also been found \cite{song91,natowitz02b}. These differ from the correlations presented here in that they are found after asymmetry, surface and Coulomb effects are taken into account. 

A more detailed analysis of the self-consistent mean-field data is presented in Table~\ref{table}, where the saturation and critical properties of the effective forces considered are presented in decreasing order of their critical temperature. The largest critical temperatures correspond to forces which have at the same time a relatively large compressibility and $\frac{m^*}{m} \sim 1$. Both inputs are therefore potential candidates for physically relevant quantities in a reliable determination of the critical temperature. Apart from this, it is hard to find any correlation in the data presented in the table. The critical compressibility factor, $\gamma_c$, is relatively stable to the change of mean-field, generally of the order $\gamma_c \sim 0.28-0.3$. These values are well below the van der Waals prediction, but in agreement with experimental data obtained from multi-fragmentation reactions \cite{elliott02} and with more sophisticated many-body approaches \cite{rios08,soma09}. 

Gogny forces are shown separately in the lower end of the Table. Interactions with apparently very similar saturation properties, such as D250 and D1P, lead to critical temperatures with large differences. This can be taken as an indication that the difference in the momentum dependence (\emph{i.e.} in their effective masses) is relevant for critical liquid-gas phenomena. It is also interesting to note, by comparing sets D250 to D300, that there is no monotonous dependence with the compressibility. 

\begin{figure}[t]
	\begin{center}
	\includegraphics[width=0.4\linewidth]{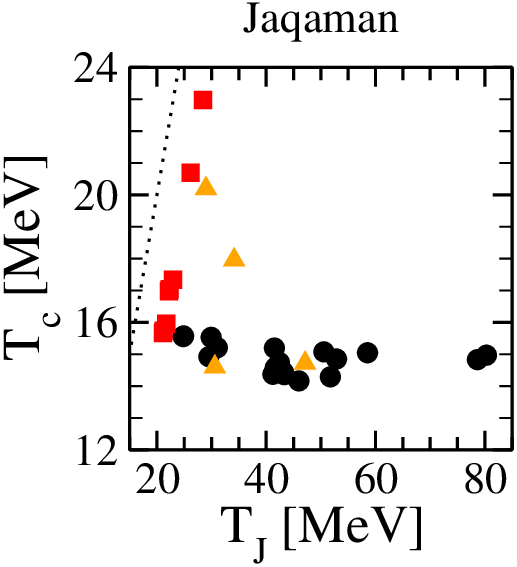}
	\caption{Self-consistent Skyrme mean-field critical temperature versus the critical temperature predicted by the Jaqaman model.}
	\label{fig:jaqaman}
	\end{center}
\end{figure}

\begin{figure}[t]
	\begin{center}
	\includegraphics[width=\linewidth]{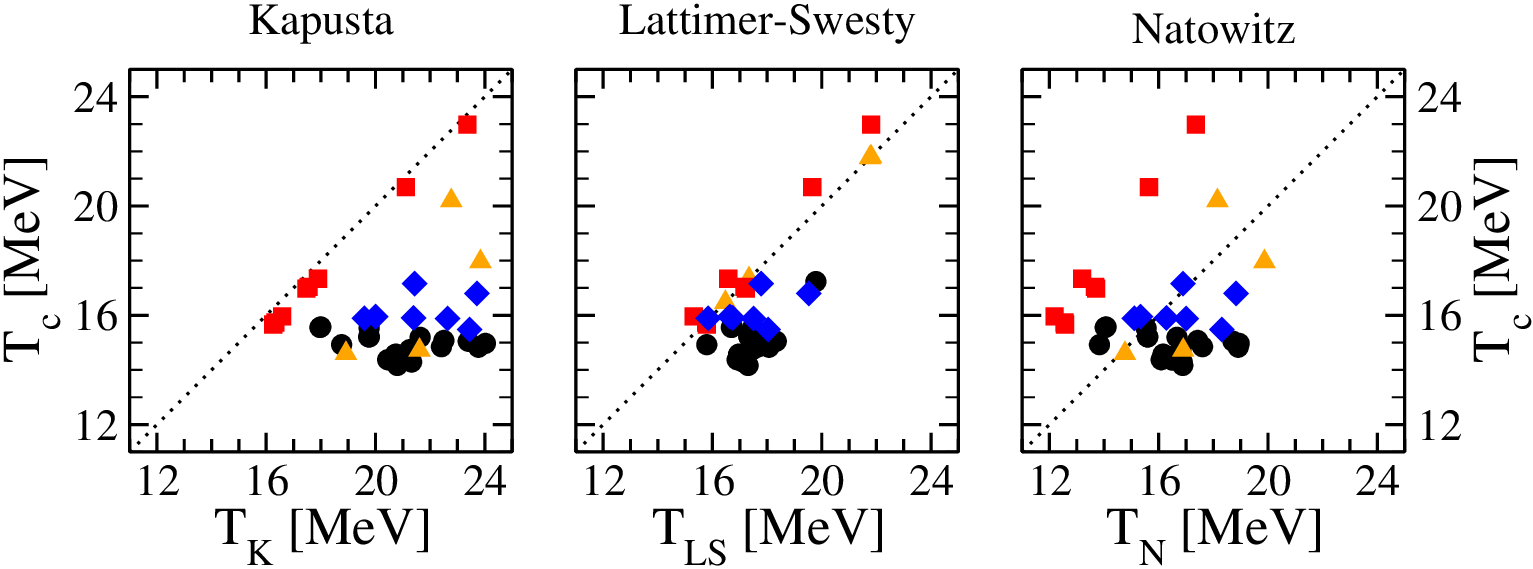}
	\caption{Self-consistent mean-field critical temperature versus the critical temperature predictions of the Kapusta (left panel), Lattimer-Swesty (central panel) and Natowitz (right panel) models. }
	\label{fig:modelstc}
	\end{center}
\end{figure}

\begin{figure}[t]
	\begin{center}
	\includegraphics[width=0.4\linewidth]{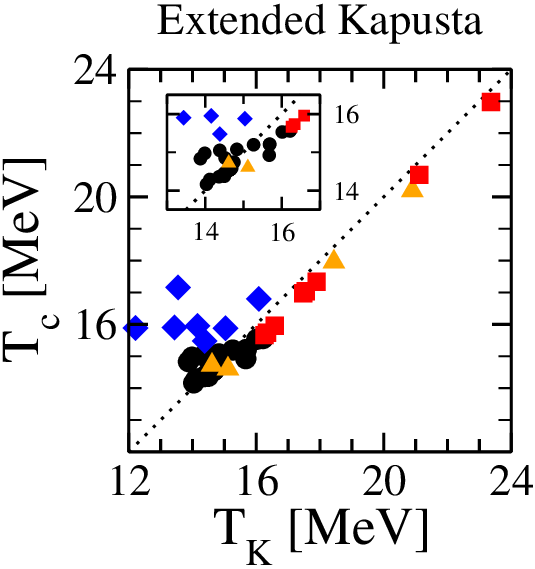}
	\caption{Self-consistent mean-field critical temperature versus the critical temperature predictions of the extended Kapusta model. The inset focuses on the region of low critical temperatures. }
	\label{fig:extkap}
	\end{center}
\end{figure}

A comparison between these critical temperatures and those predicted by the models presented in the previous Section should be helpful in establishing which are the relevant ingredients in the determination of the critical properties. A numerical experiment has been performed in which the predictions of the critical temperature for each model have been obtained from the saturation properties of the Skyrme and Gogny forces and compared to the $T_c$ obtained with the self-consistent mean-field data. The results of these experiments are summarized in Figs.~\ref{fig:jaqaman}, \ref{fig:modelstc} and \ref{fig:extkap}, which show the ``real" $T_c$ against the predictions of the different models. Good predictions should line up close to the diagonal, which is highlighted with a dotted line.

\pagebreak
\begin{longtable}[t!]{l|cccc|cccc}
$\phantom{a}$  &  \multirow{2}{*}{} $\rho_0$ & $e_0$ & $K_0$ & $\frac{m^*_0}{m}$& $ \rho_c $ & $T_c$ & $ \gamma_c $ \\
Force & [fm$^{-3}$] & [MeV] & [MeV] & $\phantom{a}$ & [fm$^{-3}$] & [MeV] &  \\
\hline
 ZR1     & $   0.173$ & $  -16.98$ & $   398.7$ & $   1.000$ & $   0.068$ & $   22.98$ & $   0.333$ \\
 ZR2     & $   0.173$ & $  -16.99$ & $   324.8$ & $   1.000$ & $   0.064$ & $   20.69$ & $   0.312$ \\
 SI      & $   0.155$ & $  -15.99$ & $   370.4$ & $   0.911$ & $   0.060$ & $   20.20$ & $   0.333$ \\
 SIII    & $   0.145$ & $  -15.85$ & $   355.4$ & $   0.763$ & $   0.056$ & $   17.96$ &  $   0.334$ \\
 SJ1     & $   0.175$ & $  -16.99$ & $   232.1$ & $   1.000$ & $   0.058$ & $   17.34$ & $   0.277$ \\
 SV      & $   0.155$ & $  -16.05$ & $   305.7$ & $   0.383$ & $   0.070$ & $   17.24$ & $   0.379$ \\
 T6      & $   0.161$ & $  -15.96$ & $   235.9$ & $   1.000$ & $   0.054$ & $   17.04$ & $   0.285$ \\
 SkT4    & $   0.159$ & $  -15.95$ & $   235.5$ & $   1.000$ & $   0.054$ & $   16.98$ & $   0.285$ \\
 ZR3     & $   0.175$ & $  -16.99$ & $   198.8$ & $   1.000$ & $   0.055$ & $   15.96$ & $   0.261$ \\
 SkT5    & $   0.164$ & $  -16.00$ & $   201.7$ & $   1.000$ & $   0.053$ & $   15.74$ & $   0.269$ \\
 SkP     & $   0.162$ & $  -15.95$ & $   201.0$ & $   1.000$ & $   0.052$ & $   15.67$ & $   0.269$ \\
 SkO     & $   0.160$ & $  -15.78$ & $   222.8$ & $   0.894$ & $   0.052$ & $   15.57$ & $   0.280$ \\
 SkOp    & $   0.160$ & $  -15.73$ & $   222.1$ & $   0.896$ & $   0.052$ & $   15.56$ & $   0.280$ \\
 BSk17   & $   0.159$ & $  -16.05$ & $   241.7$ & $   0.800$ & $   0.053$ & $   15.53$ & $   0.287$ \\
 Gs      & $   0.158$ & $  -15.59$ & $   237.3$ & $   0.784$ & $   0.053$ & $   15.21$ & $   0.288$ \\
 Rs      & $   0.158$ & $  -15.59$ & $   237.4$ & $   0.783$ & $   0.053$ & $   15.21$ & $   0.288$ \\
 SkI1    & $   0.163$ & $  -16.26$ & $   246.7$ & $   0.696$ & $   0.056$ & $   15.20$ & $   0.289$ \\
 SkI4    & $   0.162$ & $  -16.16$ & $   250.8$ & $   0.650$ & $   0.057$ & $   15.08$ & $   0.292$ \\
 SGI     & $   0.154$ & $  -15.89$ & $   261.7$ & $   0.608$ & $   0.056$ & $   15.05$ & $   0.303$ \\
 SkI3    & $   0.158$ & $  -15.99$ & $   258.3$ & $   0.577$ & $   0.058$ & $   14.97$ & $   0.302$ \\
 LNS     & $   0.175$ & $  -15.31$ & $   210.8$ & $   0.826$ & $   0.057$ & $   14.92$ & $   0.275$ \\
 SkI6    & $   0.159$ & $  -15.88$ & $   248.2$ & $   0.640$ & $   0.056$ & $   14.85$ & $   0.294$ \\
 SkI5    & $   0.156$ & $  -15.84$ & $   255.7$ & $   0.579$ & $   0.057$ & $   14.83$ & $   0.302$ \\
 SkI2    & $   0.157$ & $  -15.73$ & $   240.4$ & $   0.685$ & $   0.054$ & $   14.74$ & $   0.289$ \\
 RATP    & $   0.160$ & $  -16.05$ & $   239.5$ & $   0.667$ & $   0.055$ & $   14.72$ & $   0.287$ \\
 SkM*    & $   0.160$ & $  -15.77$ & $   216.6$ & $   0.789$ & $   0.052$ & $   14.61$ & $   0.276$ \\
 SLy0    & $   0.161$ & $  -16.02$ & $   230.2$ & $   0.698$ & $   0.054$ & $   14.58$ & $   0.281$ \\
 SLy1    & $   0.160$ & $  -15.98$ & $   229.8$ & $   0.698$ & $   0.054$ & $   14.55$ & $   0.282$ \\
 SLy3    & $   0.160$ & $  -15.97$ & $   229.9$ & $   0.696$ & $   0.054$ & $   14.55$ & $   0.282$ \\
 SLy5    & $   0.160$ & $  -15.98$ & $   229.9$ & $   0.697$ & $   0.054$ & $   14.55$ & $   0.282$ \\
 SLy230a & $   0.160$ & $  -15.99$ & $   229.9$ & $   0.697$ & $   0.054$ & $   14.54$ & $   0.282$ \\
 SLy8    & $   0.160$ & $  -15.97$ & $   229.9$ & $   0.696$ & $   0.054$ & $   14.54$ & $   0.282$ \\
 SLy4    & $   0.160$ & $  -15.97$ & $   229.9$ & $   0.695$ & $   0.054$ & $   14.52$ & $   0.282$ \\
 SLy6    & $   0.159$ & $  -15.92$ & $   229.8$ & $   0.690$ & $   0.054$ & $   14.48$ & $   0.282$ \\
 SLy7    & $   0.158$ & $  -15.89$ & $   229.7$ & $   0.688$ & $   0.054$ & $   14.44$ & $   0.282$ \\
 SLy2    & $   0.158$ & $  -15.71$ & $   226.2$ & $   0.699$ & $   0.053$ & $   14.37$ & $   0.282$ \\
 SkMP    & $   0.157$ & $  -15.57$ & $   231.0$ & $   0.654$ & $   0.054$ & $   14.29$ & $   0.286$ \\
 SLy9    & $   0.151$ & $  -15.79$ & $   229.8$ & $   0.666$ & $   0.052$ & $   14.16$ & $   0.284$ \\
\hline
D250 & $   0.158$ & $  -15.84$ & $   249.9$ & $   0.702$ & $   0.061$ & $   17.16$ & $   0.316$ \\
D300 & $   0.156$ & $  -16.22$ & $   299.1$ & $   0.615$ & $   0.058$ & $   16.80$ & $   0.318$ \\
D1M  & $   0.165$ & $  -16.02$ & $   225.0$ & $   0.746$ & $   0.058$ & $   15.95$ & $   0.292$ \\
D1   & $   0.166$ & $  -16.30$ & $   229.4$ & $   0.670$ & $   0.060$ & $   15.90$ & $   0.297$ \\
D1S  & $   0.163$ & $  -16.01$ & $   202.9$ & $   0.697$ & $   0.060$ & $   15.89$ & $   0.296$ \\
D1P  & $   0.170$ & $  -15.25$ & $   254.1$ & $   0.671$ & $   0.063$ & $   15.88$ & $   0.306$ \\
D260 & $   0.160$ & $  -16.25$ & $   259.5$ & $   0.615$ & $   0.059$ & $   15.48$ & $   0.301$ \\
	\caption{Saturation (columns 2 to 5) and critical properties (columns 6 to 8) for different Skyrme and Gogny interactions in decreasing order of $T_c$. }
	\label{table}
\end{longtable}

Fig.~\ref{fig:jaqaman} focuses on the difference between the critical temperature obtained with the Jaqaman model, Eq.~(\ref{eq:jaqaman_tc}), and that of the self-consistent calculations. A direct comparison with the pressure of Eq.~(\ref{eq:jaqaman}) can only be established for Skyrme mean-fields. More specifically, in Fig.~\ref{fig:jaqaman} the critical point obtained in Skyrme self-consistent calculations has been compared to those obtained by using the Jaqaman predictions with the corresponding values of $t_0, t_3$ and $\alpha$. The Jaqaman critical temperature is always larger than that of the self-consistent mean-field model. For effective interactions with $m^*=m$, the self-consistent $T_c$ follows the trend of the Jaqaman model critical temperature, but it is always shifted upwards by about $5-6$ MeV. Note that, for these interactions, the pressure is not influenced by the momentum dependent terms of the mean-field and, in the semiclassical limit, will presumably be well described by Eq.~(\ref{eq:jaqaman}). For the mean-fields with momentum dependence, however, the discrepancies are quite larger. The closest prediction is $8$ MeV off, and some differences are as large as $80$ MeV. Such a large discrepancy indicates that this model is unable to reproduce the thermodynamical properties of the self-consistent case. Extensions of this model to include effective mass effects as well as degeneracy corrections were already discussed in Ref.~\cite{jaqaman83} and might presumably improve the agreement between the different predictions of $T_c$. 

The results for the original Kapusta model are shown in the left panel of Fig.~\ref{fig:modelstc}. The predictions for the subgroup of Skyrme forces with $m^*=m$ (squares) are extremely good, falling very close to the diagonal. For the remaining subsets of Skyrme and Gogny forces, Eq.~(\ref{eq:kapusta_tc}) produces substantially worse results and overestimates the critical temperature by up to $9$ MeV. The Kapusta prediction for the critical temperature is therefore valid for mean-fields which have an almost quadratic and temperature independent momentum dependence. For Skyrme forces, the terms proportional to $t_1, t_2$ and $x_2$ determine, among other things, the effective mass and are presumably responsible for the worsening observed when $m^*_0 \neq m$. 

As discussed in the previous Section, the effective mass is accounted for in a very basic way in the Kapusta model. The differences between the different subgroups of Skyrme forces, with and without effective masses, highlight its importance in hot nuclear systems. In particular, the effective mass has a double effect on the thermodynamical properties. On the one hand, it influences their density dependence, as can be seen by comparing Eqs.~(\ref{eq:kapusta_pres}) and (\ref{eq:exkap_pres}). On the other, for a fixed density, the effective mass also affects the temperature dependence, \emph{i.e.} the variations of thermodynamical properties with temperature are different for different effective masses. Since the critical properties are determined by the density derivatives and the temperature dependence of the equation of state, they are particularly sensitive to the density dependence of $m^*$.

The Lattimer-Swesty and Natowitz predictions, Eqs.~(\ref{eq:lattimer_t}) and (\ref{eq:natowitz_t}), have been obtained from fits to sets of theoretical predictions to the critical temperature. Their accuracy will therefore be limited by the original data used to determine the coefficients $C_{LS}$ and $C_{N}$. One might, however, try to push these parameterizations beyond their original purpose and compare their predictions to the mean-field critical points. The results of such analysis are presented in the central and right panels of Fig.~\ref{fig:modelstc}. On average, it appears that the Lattimer-Swesty predictions are better (closer to the diagonal), than the Natowitz predictions are. For any given interaction, the difference between the two predictions is caused by (a) the difference in the coefficients and (b) the effective mass at saturation, $m^*_0$. For the Skyrme forces with $m^*=m$, the latter factor is not present and, since $C_N < C_{LS}$, the predictions for the critical temperature of the Natowitz model underestimate the critical temperature more than the Lattimer-Swesty predictions do. For the Skyrme mean-fields with effective masses, the Lattimer-Swesty and the Natowitz predictions represent an improvement with respect to the Kapusta model. All the Lattimer-Swesty results fall below the diagonal, but they overestimate the critical temperature only by about $2$ MeV on average (to be compared to almost $5$ MeV for the Kapusta model). The Natowitz formula yields predictions which are equally distributed at either side of the diagonal, but with a slightly larger deviation from the central results. By looking at these figures, it is hard to decide which of the two models produces a better agreement with the mean-field predictions.

The extended Kapusta model yields a qualitatively better description of the critical temperature in terms of saturation properties. In Fig.~\ref{fig:extkap}, the results of Eq.~(\ref{eq:exkapustaeff}) are compared to mean-field critical temperatures. For all the Skyrme forces considered, the agreement between the predicted and the mean-field $T_c$ is almost perfect, good to less than $1$ MeV accuracy. Note that for the Skyrme forces with $m^*=m$, the extended model predictions are the same as those predicted by the original Kapusta model. For the forces that have a density-dependent effective mass, the extended model supposes a substantial improvement with respect to any of the previously discussed models. For the forces with $T_c > 16$ MeV, in particular, the predictions follow closely the Skyrme mean-field results. Below this critical temperature, the extended Kapusta results show a slight tendency to underestimate the critical temperature (see inset), but the results remain always close to the diagonal. Comparing these results to other models, it is clear that the density dependence of the effective mass is a capital ingredient in the determination of the critical temperature. 

The predictions of the extended Kapusta model are worse for the Gogny interaction. The maximum observed deviation is of $\sim 4$ MeV, much larger than the largest deviation observed for the Skyrme mean-fields. These differences, however, are similar to those of the Lattimer-Swesty and Natowitz predictions for Gogny forces. The sources of the discrepancy might be attributed to several factors. To begin with, the density dependence of the effective mass for Gogny interactions is not given by that of Eq.~(\ref{eq:effmass}). Moreover, the more realistic momentum dependence introduce modifications in the temperature dependence of the pressure that might not be captured properly in the model. These considerations challenge the validity of Eq.~(\ref{eq:exkap_pres}). On this account, the extended Kapusta model might need to be reconsidered for mean-fields other than those of the Skyrme type. 

In line with the findings of Ref.~\cite{natowitz02b}, the agreement between the extended Kapusta model and the mean-field data suggests that an accurate determination of nuclear limiting temperatures might be used to derive saturation properties of nuclear matter and, perhaps, identify the density dependence of the effective mass. No matter what theoretical framework is chosen to describe the phase transition, nuclear mean-field dependences should be pinned down carefully. Universal behaviors, such as the connection between $T_f$ and $T_c$ observed here, should be identified with a comprehensive set of effective interactions. A proper treatment of shell, deformation and continuum contributions at finite temperature also seems to be a necessary ingredient for successful descriptions of data. More importantly, a treatment of clusterization and thermal fluctuations beyond the oversimplified mean-field picture is necessary to provide a realistic description of nuclear matter close to a phase transition. 

\section{Critical exponents}
\label{sec:critical}
Critical exponents characterize the properties of phase transitions in a universal way \cite{huang,pathria,fisher74,kadanoff67}. In the nuclear physics context, they have been introduced to analyze yields of multi-fragmentation reactions and the onset of percolation \cite{elliott02,kleine01}. The critical exponents of infinite nuclear matter have received far less attention. Calculations can easily be performed in simplified models, as those presented in Section \ref{sec:models}, but they are harder to implement numerically for self-consistent mean-field models. Some critical exponents have been computed for a single Skyrme mean-field in Ref.~\cite{ducoin06b} to address issues related to bimodality in isospin asymmetric matter. 

The calculations of critical exponents for different mean-fields are of interest for theoretical reasons. On the one hand, simplified models, as has been discussed in the two previous sections, might not be able to reproduce the liquid-gas phase transition properties of mean-field (or more complicated many-body) calculations. Consequently, their predictions for critical exponents might not be reliable. Extensive mean-field calculations are needed to confirm numerically the values of the exponents and to check its independence from the underlying effective interaction. It is particularly interesting to note that the self-consistency at the mean-field level can only be imposed numerically and thus precludes any analytical development. On the other hand, it is well-known from magnetic systems that critical exponents are sensitive to the range of the interaction \cite{fisher74}. Therefore, it is plausible that self-consistent mean-field calculations of homogeneous nuclear matter have different critical exponents for interactions which have either zero or finite range. 

Critical exponents give unique information on the behavior of the thermodynamical properties of nuclear matter close to the critical point. A numerical code has been implemented to compute three of these exponents for mean-field calculations of nuclear matter. The order parameter in the liquid-gas phase transition is given, at each temperature, by the difference of the liquid and the gas phase densities, \emph{i.e.}~the coexistence line. The first exponent, $\beta$, characterizes how the order parameter approaches the critical point\footnotemark[1]:
\begin{align}
	\rho_l - \rho_g \sim (-\tau) ^\beta \, ,
	\label{eq:beta}
\end{align}
where the reduced temperature, $\tau=\frac{T-T_c}{T_c}$, is a measure of the distance to the critical point. The critical exponent $\gamma$ is given by the isothermal compressibility,
\begin{align}
	\frac{1}{K_T} =\rho \left. \frac{\partial p}{\partial \rho} \right|_T \, ,
	\label{eq:compr}
\end{align}
close to the critical temperature\footnote[1]{Strictly speaking, one should consider different critical exponents for the liquid and gas branches at $\tau<0$, so that ($\beta, \gamma, \delta$) would become ($\beta_l, \gamma_l, \delta_l$) and ($\beta_g, \gamma_g, \delta_g$). The numerical results presented below fulfill $\beta_g=\beta_l$, $\gamma_g=\gamma_l$ and $\delta_g=\delta_l$ to within $5\%$, $6\%$ and $10\%$ accuracy, respectively. The results that are quoted have been obtained by averaging liquid and gas branches to eliminate uncertainties in the determination of $\rho_c$. Moreover, for $\tau>0$ and $\rho=\rho_c$ one can also define a $\gamma$ critical exponent \cite{kadanoff67}. Scaling laws suggest $\gamma=\gamma'$, which is fulfilled numerically to a $1 \%$ accuracy.},
\begin{align}
	K_T \sim (-\tau) ^{-\gamma} \, .
	\label{eq:gamma}
\end{align}
Moreover, the density dependence of the critical isotherm is described by the critical exponent, $\delta$\footnotemark[1]:
\begin{align}
	| p - p_c | \sim | \rho - \rho_c |^\delta  \, .
	\label{eq:delta}
\end{align}
Finally, for nuclear matter, the heat capacity at $\rho=\rho_c$ also approaches the critical point with a power-law behavior:
\begin{align}
	c_V = T \frac{\partial S}{\partial T} \sim \tau^{-\alpha} \, .
	\label{eq:alpha}
\end{align}
$\alpha$ will not be explicitly computed here, but its value could be derived from the other critical exponents by means of either of the following scaling law relations \cite{kadanoff67}:
\begin{align}
	2 - \alpha = \beta(\delta+1) , \qquad 2-\alpha=\gamma + 2 \beta \, .
	\label{eq:scaling}
\end{align}

The critical exponent $\beta$ is closely related to the temperature dependence of the coexistence curve. A basic property of the liquid-gas phase transition for atomic substances is the law of corresponding states, which dictates that the equation of state of simple atomic liquids and gases is the same after normalizing it to the respective critical properties \cite{huang, guggenheim45}. An important consequence of this law is that, after rescaling by the critical density and temperature, the coexistence curve of a wide range of substances is the same. It is important to note that this is an empirical law, which indicates that the thermodynamics of all these substances is similar. In the nuclear matter case, one might wonder whether different coexistence lines, obtained from different mean-fields, follow a similar law of corresponding states. Analogous studies have been carried out in the context of molecular dynamics simulations of classical liquids \cite{okumura00}. For that particular case, the coexistence curve obtained with different inter-atomic potentials fulfills a law of corresponding states.

\begin{figure}[t]
	\begin{center}
	\includegraphics[width=0.55\linewidth]{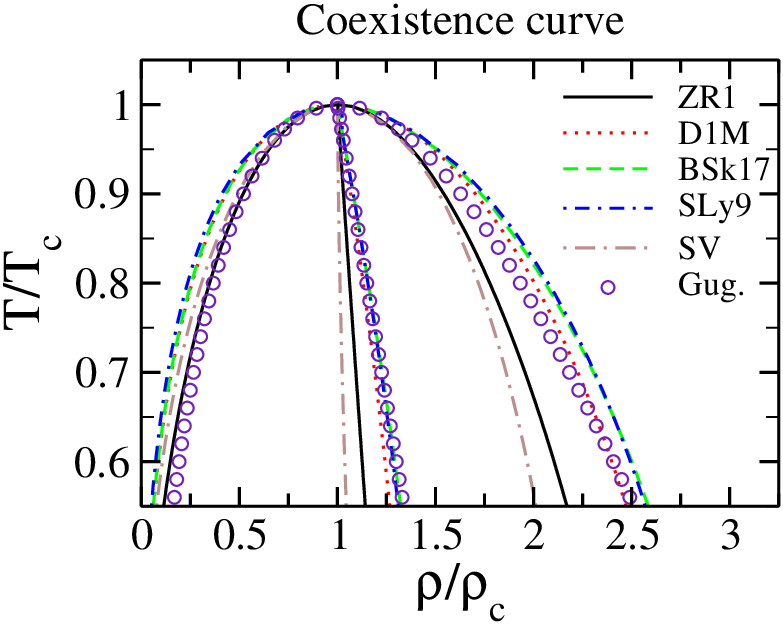}
	\caption{Normalized coexistence curve for different mean-fields. The small circles correspond to the Guggenheim parameterization, Eq.~(\ref{eq:gug}). The diameters of each mean-field are also shown.}
	\label{fig:norm_coex}
	\end{center}
\end{figure}

The scaled coexistence curves for a few representative mean-fields are shown in Fig.~\ref{fig:norm_coex}. SV (long-dash-dotted line) and SLy9 (short-dash-dotted line) represent extreme behaviors of the coexistence curves. The remaining mean-field coexistence curves fall within these two results. While in the gas phase there seems to be a certain degree of agreement between different mean-fields, substantial differences are found in the liquid branch even for values of $T/T_c \sim 0.9$, relatively close to the critical point. The breakdown of the law of corresponding states provides an interesting insight on the interaction dependence of the thermodynamical properties. To begin with, it indicates that the equation of state of different mean-fields is not simply related to one another by a simple scaling with the critical properties. This is in contrast to classical liquids, in which plausibly the differences in interactions are less acute and such a scaling is accurate \cite{okumura00}. As a matter of fact, on physical grounds, it is only the thermodynamics close to the critical point that should be interaction independent and dictated by the critical exponents. The mean-field predictions of Fig.~\ref{fig:norm_coex} are compared to the Guggenheim parameterization:
\begin{align}
\frac{\rho_{l,g}}{\rho_c} = 1 - \frac{3}{4} \tau \pm \frac{7}{4} (-\tau)^\beta \, ,
\label{eq:gug}
\end{align}
where the $+$ ($-$) sign is used for the liquid (gas) branch \cite{guggenheim45}. Instead of the original $\beta=\frac{1}{3}$, which is suitable for classical gases \cite{kadanoff67}, the value $\beta=\frac{1}{2}$ has been used, in accordance to the results discussed in the following paragraphs. The Guggenheim formula produces a somewhat average behavior, falling within all the mean-field results and might be used for phenomenological applications. 

Useful information regarding the phase coexistence can also be gathered by analyzing the so-called ``diameter" of the transition, \emph{i.e.}~the average of the coexistence densities at each temperature. The diameter is given by the almost vertical, central line shown in Fig.~\ref{fig:norm_coex}. The tilt of the diameter with respect to the vertical indicates that the coexistence curve is not perfectly symmetric. As a matter of fact, the law of rectilinear diameter sates that the diameter scales linearly with $\tau$:
\begin{align}
\frac{\rho_{g}+\rho_l}{2\rho_c} =1 - c \tau \, ,
\label{eq:rectilinear}
\end{align}
where $c$ is some numerical coefficient \cite{okumura00,kadanoff67,guggenheim45}. As observed in Fig.~\ref{fig:norm_coex}, the diameter is indeed rectilinear for all the mean-fields considered within a wide range of temperatures (the linear behavior breaks down very close to $T_c$, with a deviation related to $\alpha$ \cite{mermin71}). Every mean-field predicts a different tilt (\emph{i.e.}~a different value of $c$) of the diameter. Some of these agree with the value $c = \frac{3}{4}$, predicted by Eq.~(\ref{eq:gug}), but generally lower values (as low as $c \sim 0.13$) are found. These indicate that mean-field coexistence curves are somewhat more symmetric than the Guggenheim parameterization. 

\begin{figure}[t]
	\begin{center}
	\includegraphics[width=0.55\linewidth]{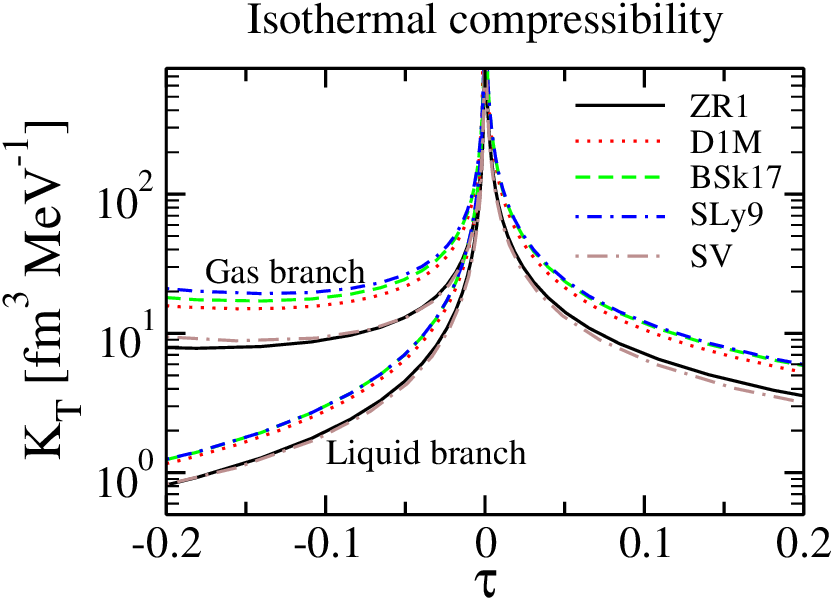}
	\caption{Isothermal compressibility as a function of reduced temperature for different mean-fields. Below $T_c$, the liquid and gas branches are obtained at the corresponding coexistence densities. Above $T_c$, $K_T$ is computed at $\rho_c$. }
	\label{fig:compt}
	\end{center}
\end{figure}

The isothermal compressibility diverges close to the critical point with a power law behavior governed by the $\gamma$ critical exponent. $K_T$ is shown in Fig.~\ref{fig:compt} as a function of the reduced temperature for different mean-fields. Below $T_c$, the gas and the liquid branch are shown independently. A fairly large variation of $K_T$ is observed for the different mean-fields in the gas region (note the logarithmic scale), in contrast to the relatively similar gas coexistence curves seen in Fig.~\ref{fig:norm_coex}. For the liquid branch, the predictions for $K_T$ are relatively closer to each other. It is important to note that $K_T$ is not related to the nuclear compressibility, $K_0$. Only in the $T \to 0$ limit, the value of $K_T$ in the liquid branch is given by the compressibility and the saturation density, $K_T = \frac{9}{K_0 \rho_0}$. Forces with larger compressibilities, such as ZR1 and SV, will therefore lead generally to lower $K_T$'s. Above the critical point, $K_T$ is taken at $\rho=\rho_c$ and its value decreases steadily, in accordance to the $K_T \sim \tau^{-1}$ behavior. Again, a certain degree of spread between different mean-field predictions is found, particularly among those with different nuclear compressibilities. Note that, in spite of these differences, the critical exponent $\gamma$ is the same for all the interactions considered  (see below).

\begin{figure}[t]
	\begin{center}
	\includegraphics[width=0.75\linewidth]{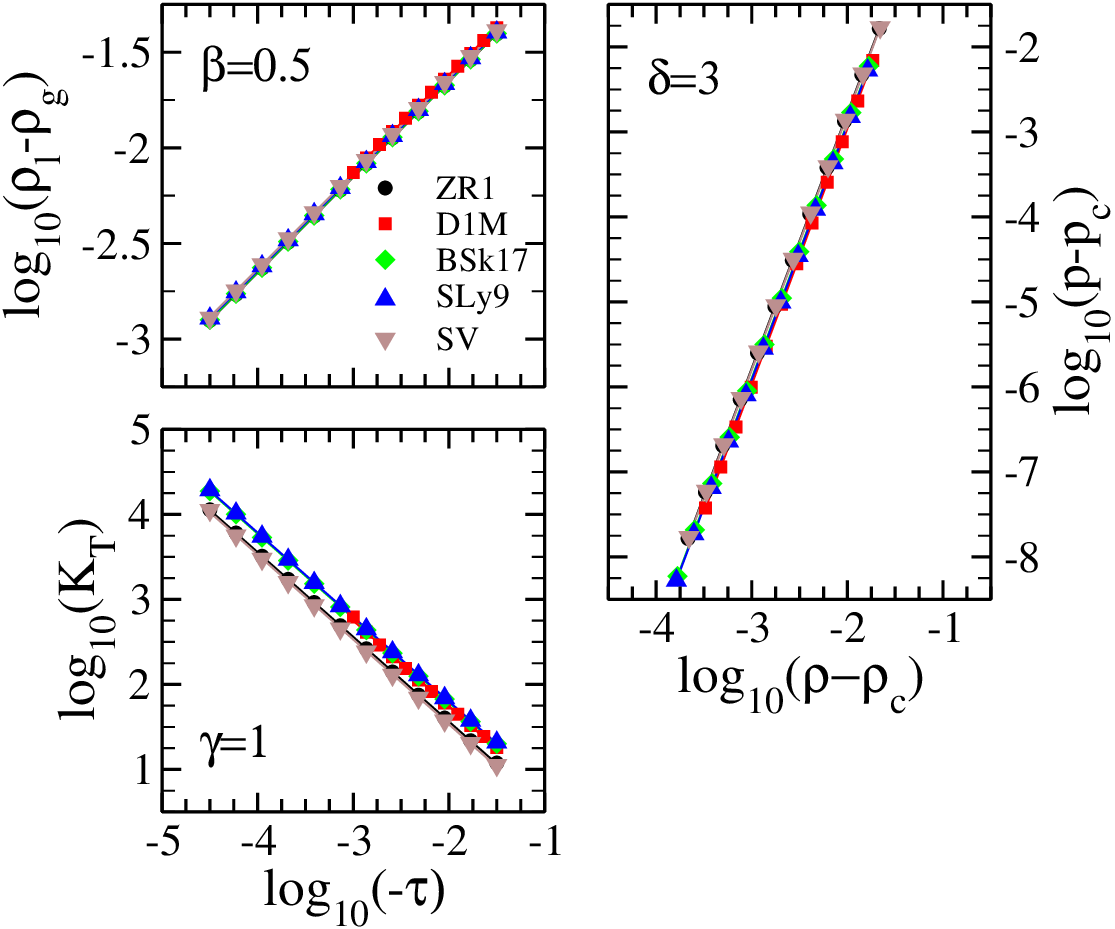}
	\caption{Logarithmic plots used to determine the critical exponents. Left upper panel: difference between liquid and gas coexistence densities versus reduced temperatures. Left lower panel: isothermal compressibility versus reduced temperatures. Right panel: critical isotherm versus density. Symbols correspond to mean-field data, while lines are the results of linear regressions.}
	\label{fig:cexps}
	\end{center}
\end{figure}

For systems with no thermal fluctuations, the Landau mean-field theory of phase transition predicts the following values for the critical exponents \cite{kadanoff67}:
\begin{align}
	\beta=\frac{1}{2}, \quad \gamma=1, \quad \delta=3 \, , \quad \alpha=0 \, .
	\label{eq:landau}
\end{align}
A first indication of the nuclear matter critical exponents can be obtained by the models introduced in Section \ref{sec:models}. Indeed, the Jaqaman model reproduces the Landau theory results \cite{jaqaman83,goodman84} and one can easily show that the Kapusta model (and its extension) also fall in the same category. As mentioned earlier, however, these basic parameterizations of the equation of state are too simple to describe the liquid-gas coexistence region of nuclear matter derived from self-consistent calculations. The latter provide a consistent description of hot nuclear matter, with no assumptions whatsoever on temperature or density dependences, providing a more sound analysis from a theoretical perspective. It is clear, however, that the mean-field description has important limitations in describing nuclear liquid-gas phenomena. As a consequence, the following results concerning critical exponents should not be taken as realistic, but rather as a theoretical confirmation of the validity of Landau mean-field theory for homogeneous nuclear matter.

The self-consistent calculations give critical exponents of the Landau mean-field type. As expected from physical grounds, this result is independent of the effective interaction under consideration. To prove the latter point, Fig.~\ref{fig:cexps} shows the logarithmic plots necessary to determine the values of $\beta$, $\gamma$ and $\delta$. The data displays a perfectly linear behavior. The slopes of these lines correspond to the critical exponents, and they are indeed independent of the interaction. In agreement with the previous considerations, however, the offsets of the lines do depend to a certain extent on the interaction. 

It is important to indicate that, for an accurate calculation of these exponents, the coexistence line needs to be determined reliably for temperatures very close to the critical point. In addition, both the critical temperature and density need to be known with relative precisions of at least 4 significant digits. While this is of a lesser concern for Skyrme forces, the calculations for finite-range Gogny mean-fields have proven to be harder and involve longer computational times. For these interactions, coexistence curves have been determined typically down to  $\tau \sim 10^{-3}$ and, as a consequence, somewhat less accurate results have been obtained. Note that the data for $K_T$ has been obtained as an average of the gas and the liquid branches. Similarly, for the critical isotherm, densities above and below $\rho_c$ have been averaged. This procedure cancels out part of the numerical uncertainties in the determination of the critical density and temperature.

\begin{table}[t!]
\begin{center}
\begin{tabular}{l | c c c}
$\phantom{a}$ & $\beta$ & $\gamma$ & $\delta$ \\
\hline
ZR1 & 	0.4996$\pm$0.0001 &     	0.9951$\pm$0.0013 	& 3.0000$\pm$0.0001 \\
D1M &	0.5019$\pm$0.0017 & 	1.0074$\pm$0.0103 	& 3.0124$\pm$0.0034 \\
BSk17 & 	0.4993$\pm$0.0002 &  	0.9934$\pm$0.0017 	& 3.0006$\pm$0.0002 \\
SLy9 & 	0.4989$\pm$0.0003 &	0.9926$\pm$0.0020 	& 3.0007$\pm$0.0002 \\
SV & 	0.5001$\pm$0.0006 &	0.9978$\pm$0.0036 	& 3.0067$\pm$0.0005 \\
\hline
\end{tabular} 
\end{center}
\caption{Critical exponents obtained from linear regression fits of the results of Fig.~\ref{fig:cexps}. Quoted errors correspond to the standard deviations of the slope.}
\label{table:cexps}
\end{table}

The critical exponents have been obtained numerically by a linear regression of the mean-field data presented in Fig.~\ref{fig:cexps}. To determine $\beta$ and $\gamma$, a logarithmic mesh of 12 values between $\tau = 10^{-1.5}$ and $10^{-4.5}$ ($10^{-3}$)  has been used for Skyrme (Gogny) forces. Similarly, to find $\delta$, 12 logarithmic points have been distributed between $\frac{\rho-\rho_c}{\rho_c} = 10^{-0.5}$ and $10^{-2.25}$. The results of the slopes of the corresponding regressions for the five interactions considered in this Section are presented in Table~\ref{table:cexps}. The quoted errors correspond to the standard deviations of the slope and can be taken as a measure of the quality of the fits. The extremely small values of these errors provide an indication of the very good linear behavior of the data in the logarithmic scales and therefore validate the power law behavior of the self-consistent mean-field data. The agreement with the Landau predictions is excellent, accurate to the third digit in most cases. Minute deviations from this behavior should plausibly be attributed to numerical errors associated with the limited accuracy of the self-consistent numerics close to the critical point. While only 5 mean-fields are shown here for practical purposes, identical results have been obtained for the remaining 36 Skyrme and 7 Gogny interactions considered in the previous Section. 

For real physical systems, the values of the critical exponents can deviate substantially from the Landau mean-field predictions \cite{pathria,kadanoff67}. Yields in multifragmentation reactions, for instance, can be described with $\beta \sim 0.3$ and $\gamma \sim 1.4$  \cite{gilkes94}. The fact that Landau-type exponents are found for Hartree-Fock calculations of nuclear matter is a sign of its inability to properly describe the critical properties. The self-consistent mean-field approach represents a minimization of the thermodynamical potential, subject to the constraint that the density matrix is of the one-body type \cite{bonche84}. The physical trajectories that this minimization procedure yields are unique and have small thermal fluctuations by construction. Critical exponents different from Landau theory, however, can only be achieved if thermal fluctuations, presumably larger than those introduced by one-body approximations, are taken into account \cite{kadanoff67}. It is plausible that this can be achieved by either considering correlated many-body approaches or by including additional stochastic fluctuations on top of the mean-field \cite{juillet02}.

To summarize, the normalized equations of state, coexistence curves and thermodynamical properties of nuclear matter coming from different self-consistent mean-field calculations do not follow a law of corresponding states. At low temperatures, the thermodynamical properties are actually sensitive to details of the interaction, their ranges and structures. In spite of these differences, critical exponents are the same, independently of the mean-field. The physics close to the critical point is therefore not affected by the structure of the short-range nuclear effective interaction and, at the critical point itself, the long-range thermal fluctuations dominate the thermodynamics of nuclear matter. For self-consistent mean-field calculations of infinite matter, the numerical values of the critical exponents agree with those given by Landau mean-field theory. 

\section{Conclusions and outline}
\label{sec:conc}
Hot nuclear matter has been studied within the self-consistent Hartree-Fock approach using different mean-field interactions of the Skyrme and the Gogny types. The basic aim of this work has been to provide a systematic analysis of the liquid-gas phase transition in the mean-field approximation for nuclear matter. Calculations performed with several mean-fields are helpful in elucidating which are the physically relevant properties for the liquid-gas transition, since they provide means of eliminating the mean-field dependence of the results. Subtracting this dependence, a more systematic understanding of the properties of the liquid-gas phase transition is obtained. This procedure might also be necessary in more realistic descriptions of nuclear critical phenomena. To my knowledge, this is the first time that a generic analysis with different effective interactions within a self-consistent theory is used to this end. 

A step-by-step improvement of our understanding of the temperature dependence of nuclear properties requires a full control of symmetric nuclear matter as a first, initial step. Two points stand out in the phase diagram due to their physical relevance: the saturation point and the critical point. Liquid-gas critical points have been computed for different Skyrme and Gogny interactions and correlations with zero temperature saturation properties have been investigated. Parameter spaces for physically allowed critical densities and temperatures have been identified by using a comprehensive set of mean-field calculations. On the one hand, it is difficult to find clear connections between saturation and finite temperature properties. On the other hand, a tight linear relation has been found between the flashing and the critical density, as well as between the flashing and critical temperatures. Given the physical equivalence between the flashing temperature of nuclear matter and the limiting temperature of nuclei, this correlation might be useful in relating the nuclear matter critical point to the flashing properties of nuclei found in reactions. 

A qualitative understanding of the properties of the phase transition can be obtained by using analytical parameterizations of the equation of state. The Jaqaman and the Kapusta models represent two of these parameterizations and are among the most widely used in the literature. The critical point can be solved explicitly in these models, providing relations between the critical point and parameters related to the zero-temperature equation of state. These models also yield correlations between the flashing and the critical points, in overall agreement with those found with the mean-field results. The temperature dependence of the Jaqaman and the Kapusta models is based on degeneracy expansions, either in the semi-classical or the degenerate regimes. Unfortunately, the liquid-gas phase transition in nuclear matter occurs between these two regimes and none of these models is able to predict reliably the mean-field results of the critical point. In particular, the critical temperatures are generally overestimated by few to some tenths of MeV. Lattimer-Swesty and Natowitz \emph{et al.}~have proposed phenomenological modifications to improve the predictive power of the Kapusta model. Both formulae do a better job in predicting the critical temperature of the Skyrme and Gogny forces, but fail to give quantitatively accurate results when compared to mean-field systematics. 

A common feature of the analytical models is a very crude treatment of the effective mass, in spite of the fact that $m^*$ has been identified as a basic ingredient in the physics of hot nuclei and the liquid-gas phase transition. The Kapusta model has been extended to include the density dependence of the Skyrme effective mass. While this model cannot be solved analytically, an expansion close to the critical point yields systematic corrections to the Kapusta model predictions. At first order, while the critical density is basically unaffected, the critical temperature is modified by a multiplicative factor that is equivalent to the effective mass at a suitable sub-saturation density. This expression provides an excellent prediction of the critical temperature for all the Skyrme mean-fields, to within $1$ MeV accuracy, and represents a quantitative improvement with respect to the other models. These results illustrate the importance of  the density dependence of the effective mass, \emph{i.e.}~the momentum dependence, in mean-field models at finite temperature. The extended model works worse for Gogny forces, with a tendency to underestimate the critical temperature by up to $4$ MeV. It is plausible that a modification is needed to account for the different density dependence of the effective mass with this mean-field. 

As an additional outcome to this study, the $\beta$, $\gamma$ and $\delta$ critical exponents of the transition have been computed numerically. As expected from general physical grounds, no interaction dependence is found in the critical exponents which, even for the finite-range Gogny interaction, agree with the predictions of the Landau mean-field theory of critical phenomena to less than a $1 \%$ accuracy. The lack of effective interaction dependence of the critical exponents is in contrast to the thermodynamical properties, which do depend on the mean-field interaction. As a matter of fact, if the coexistence curve of different mean-fields is rescaled to the corresponding critical properties, a clear disagreement is found \emph{i.e.} the law of corresponding states is not valid for different nuclear mean-fields. 

Further generalizations of this work to asymmetric nuclear matter and finite nuclei would provide a consistent route to connect the effective in-medium nuclear interaction and the finite temperature properties of nuclear matter. For isospin imbalanced systems, for instance, the critical temperature might be correlated to the symmetry energy (and/or its derivatives). Studies with comprehensive mean-field sets, analogous to the analysis presented here, could easily test this idea. Similarly, one can nowadays carry out fully self-consistent mean-field calculations of semi-infinite matter at finite temperature \cite{danielewicz09}. From these studies, one could determine the temperature and isospin dependence of nuclear surface properties. At relatively low temperatures, moreover, it could be interesting to analyze the temperature dependence of nuclear structural properties. Mean-field (or, for this purpose, energy density functional) nuclear structure calculations have provided and continue to provide accurate predictions for single-particle and bulk properties in wide regions of the nuclear chart. Their extension to finite temperature yields a theoretically consistent description of hot nuclei that could, for instance, yield predictions for the isotopic dependence of limiting temperatures. A large source of uncertainty in such analysis, however, is the extremely important role played by Coulomb forces. For finite systems, this is well beyond the reach of mean-field theory and additional theoretical developments might be needed to provide a realistic description of the coexistence and spinodal regions.

\section*{Acknowledgements}

The author is thankful to Artur Polls, Denis Lacroix and Paul Stevenson for their careful reading of the manuscript and for fruitful discussions. This work has been supported by a Marie Curie Intra European Fellowship within the 7$^{th}$ Framework programme, STFC grant ST/F012012 and the U.S. National Science Foundation under Grant No. PHY-0800026. 

\bibliographystyle{elsarticle_num1a}
\bibliography{main}

\end{document}